\documentclass[12pt]{article}
\usepackage{apacite} 
\usepackage{amssymb,amsmath,amsfonts,eurosym,geometry,graphicx,caption,color,setspace,sectsty,comment,caption,natbib,pdflscape,subfigure,array,hyperref, tikz, float, here, bm,enumitem}
\usepackage{booktabs}
\usepackage{longtable}
\usepackage{graphicx}
\usepackage{footnote}
\usepackage{hyperref}
\usepackage{lscape}
\usepackage[super]{nth}
\usepackage{pdfpages}
\usepackage{tabularx}
\newcolumntype{Y}{>{\centering\arraybackslash}X}
\usepackage[bottom]{footmisc}
\usepackage[normalem]{ulem}
\useunder{\uline}{\ul}{}
\usepackage{hyperref}
\hypersetup{
    colorlinks=true,
    linkcolor=blue,
    filecolor=blue,      
    urlcolor=blue,
    citecolor=blue
}
\newcolumntype{L}[1]{>{\raggedright\let\newline\\arraybackslash\hspace{0pt}}m{#1}}
\newcolumntype{C}[1]{>{\centering\let\newline\\arraybackslash\hspace{0pt}}m{#1}}
\newcolumntype{R}[1]{>{\raggedleft\let\newline\\arraybackslash\hspace{0pt}}m{#1}}

\begin{document}

\renewcommand{\arraystretch}{0.9}
\begin{titlepage}
\title{Temperature and Mental Health: Evidence from Helpline Calls\thanks{I am thankful for helpful comments from Ludovica Gazze, Jamie Mullins, Alessandro Palma, Patrick Bigler, as well as conference and seminar participants at the AERE Summer Conference, the EAERE Annual Conference, the ESPE Annual Conference, the IAERE Annual Conference, and the VfS Environmental and Resource Economics Junior Workshop. I am grateful to my supervisor Doina Radulescu for her excellent and continuous guidance. I am particularly indebted to Ludger Storch and \textit{TelefonSeelsorge} for providing the necessary data. The author has no conflict of interest in any part and did not receive any funding for this work. All remaining errors are my own.}}
\author{Benedikt Janzen\thanks{University of Bern, KPM Center for Public Management and OCCR Oeschger Centre for Climate Change Research; Email: benedikt.janzen@kpm.unibe.ch}}
\date{First version: July 2022 \\
This version: November 2022}

\maketitle

\begin{abstract}
This paper studies the short-term effects of ambient temperature on mental health using data on nearly half a million helpline calls in Germany. Leveraging location-based routing of helpline calls and random day-to-day weather fluctuations, I find a negative effect of temperature extremes on mental health as revealed by an increase in the demand for telephone counseling services. On days with an average temperature above 25$^{\circ}$C (77$^{\circ}$F) and below 0$^{\circ}$C (32$^{\circ}$F), call volume is 3.4 and 5.1 percent higher, respectively, than on mid-temperature days. Mechanism analysis reveals pronounced adverse effects of cold temperatures on social and psychological well-being and of hot temperatures on psychological well-being and violence. More broadly, the findings of this work contribute to our understanding of how changing climatic conditions will affect population mental health and associated social costs in the near future.

\noindent     
\vspace{0in}\\
\noindent\textbf{Keywords: Climate Change, Weather, Well-being} \\
\vspace{0in}\\
\noindent\textbf{JEL Codes: I1, I31, Q5, Q54} \\

\bigskip
\end{abstract}
\setcounter{page}{0}
\thispagestyle{empty}
\end{titlepage}
\pagebreak \newpage

\section{Introduction} \label{sec:introduction} Mental disorders affect nearly one billion people worldwide, accounting for five percent of the total global burden of disease \citep{alize2022global}. The \citet{WHA2012} expects that by the end of the decade mental illness will be the leading cause of disease burden worldwide, up from 13th in 1990. For those affected, mental illnesses such as depression are associated with high individual costs, as they reduce labor force participation and entail large earning penalties \citep{butikofer2020employment, biasi2021career}. Globally, the World Economic Forum estimates that the socioeconomic cost of mental disorders was \$2.5 trillion in 2010, and it is expected to more than double by 2030 \citep{WEF2011}. Because of uncertainty in potential risk factors for human health, which include climate variability and trends \citep{mcmichael2006climate}, and given an increase in global surface temperatures through at least mid-century \citep{IPCC2021} projected costs are likely to be an underestimate.

\noindent\\
Despite their large share of the global burden of disease and in the face of climate change, there is limited causal evidence on the impact of temperature extremes on mental health \citep{berry2018case}. Where existing work has explored the causal effects of temperature on mental distress, it has focused on suicidality \citep{burke2018higher, carleton2017crop}, health care utilization \citep{mullins2019temperature}, and self-reported mental health \citep{obradovich2018empirical}. However, these measures have limitations that ultimately make it difficult to infer the impact of temperature on mental health from them alone \citep{romanello20212021, obradovich2022identifying} as they either focus on rare and extreme or clinical outcomes\footnote{Mental health stigma can discourage people from seeking professional help \citep{bharadwaj2017mental}. In addition, access to evidence-based treatment remains inadequate even in high-income countries \citep{who2004prevalence, WHA2012}.}, or represent subjective measures of mental health that might suffer from systematic bias \citep[e.g.,][]{jahedi2014advantages}.

\noindent\\
This paper studies the short-term effects of ambient temperature on mental distress using administrative records from the largest general telephone counseling service in Germany, consisting of 485,274 individual calls received at 55 sites nationwide between November 2018 and March 2020. The data contains the timestamp, processing location, call topics, and caller characteristics for each call. Telephone helplines provide free, low-threshold, anonymous\footnote{In Germany, anonymous and unobservable access to telephone counseling services is even guaranteed by federal law (§99 Abs. 2 Telecommunications Act (TKG)), which prohibits telecommunication service providers to list telephone counseling calls in call detail records.} counseling services for people with unmet mental health needs. Although not a direct measure of psychiatric disorders, the use of helpline data offers many advantages including its high spatial and temporal resolution, its ability to capture subclinical mental health problems, and its potential to account for individuals who fail to seek professional treatment and are therefore not accounted for in clinical data. Compared to self-reported mental health, calls to counseling services can be seen as revealed rather than stated mental health problems, and compared to other high-resolution measures related to mental health, such as online searches, helpline calls are more likely to indicate help-seeking rather than information-seeking. Only recently, data from counseling centers were used during the COVID-19 pandemic as a real-time proxy for monitoring population mental health \citep{brulhart2021distress, armbruster2020lost} and have been shown to be an appropriate tool for behavioral predictions of suicide rates \citep{choi2020development}.\footnote{Helpline data also has certain limitations. For example, the observed number of calls may not reflect the actual demand for telephone counseling services because of capacity constraints of counseling centers or because of people who call more than once, artificially increasing the number of answered calls (so-called 'frequent callers'). See \cite{liu2021helpline} for an excellent discussion of the benefits and shortcomings of helpline call data for monitoring population-level mental distress. As argued by \cite{brulhart2021distress}, helpline data are a supplement to, rather than a replacement for, established indicators of population-level mental health.} 

\noindent\\
I interpolate daily temperature, precipitation, sunshine duration, wind speed, relative humidity, and air pollution at each counseling center using monitor readings from the universe of German weather and air quality stations spread across the entire country. As callers are routed to the nearest crisis center based on location, I can exploit idiosyncratic day-to-day variations in ambient temperatures to determine their causal impact on mental health. I use these exogenous temperature variations to answer two main questions. First, I ask whether exposure to temperature extremes affects the incidence of mental health problems. In doing so, I assume that changes in population mental health are revealed by discrete changes in the number of calls to telephone counseling services. Second, I seek to answer why thermal stressors undermine mental health. In answering this question, I assume that the mechanism is partially revealed by the underlying concern of the call.  

\noindent\\
Two-way linear fixed effects regression, in which I account for a wide range of concurrent environmental factors, shows a statistically significant adverse effect of exposure to suboptimal temperature levels on the incidence of mental health problems, with increasing effects toward the ends of the temperature distribution. I find that on days with an average temperature above 25$^{\circ}$C and below 0$^{\circ}$C, call volume is 3.4 and 5.1 percent higher, respectively, than on days with an average temperature in the middle range of the temperature spectrum. The estimates remain robust across alternative model specifications and survive in-space as well as in-time placebo tests. Effect sizes differ substantially both by age and gender. The results indicate that only the number of calls from male help-seekers increases significantly on hot days, while there is no increase in female help-seekers. On cold days, however, the exact opposite is observed. For age-related effect heterogeneity, I document that the increase in call volume on cold days is mainly driven by older individuals and the increase in call volume on hot days is due to young individuals. 

\noindent\\
Based on the detailed call-level data, I estimate whether counseling topics change with temperature levels. Several interesting results emerge from this analysis. The findings suggest negative effects of hot temperatures on calls about psychological well-being (e.g., mood, stress, self-image) and violence (i.e., physical, mental, and sexual violence), as well as negative effects of cold temperatures on psychological and social well-being (i.e., relationships). For example, I document that the number of calls about social relations increases by 5.7 percent on extremely cold days, relative to mid-temperature days. I interpret this finding as evidence that cold temperatures have the potential to undermine mental health by decreasing social well-being, potentially as a consequence of isolation and limited mobility. At the other end of the temperature spectrum, results indicate that the number of calls about psychological well-being is 7.0 percent higher on extremely hot days than on days with average temperatures, suggesting a causal psychological link between hot temperatures and mental health. 

\noindent\\
When looking at the impact of past temperature exposure on mental health, the results provide evidence of the existence of an immediate adverse effect of hot temperatures on mental health, with no clear evidence of delayed effects. For extremely cold temperatures, my results suggest that the negative effect is caused by constant exposure over several days, further supporting the interpretation that cold temperatures primarily affect mental health through social isolation.  

\noindent\\
Additionally, I use the individual-level call data to estimate how temperature changes affect treatment resource utilization, as measured by the length of telephone counseling sessions. I find that temperature affects the intensity to which helplines are used and document that the duration of counseling monotonically decreases in temperature levels, even after controlling for compositional changes of callers. Results suggest that an increase in temperature by 1$^{\circ}$C translates to a decrease in call length by 0.3 percent.      

\noindent\\
The main contribution of this work is to use a complementary and novel measure of mental health to overcome some of the limitations of conventional measures in assessing the effects of extreme temperatures at the population level, and to provide further causal evidence of the effects of suboptimal temperature levels on mental health. Most of the empirical literature to date has focused on single weather events, a very narrow geographic area, allows only for correlational evidence, or does not consider the full distribution of temperature \citep[e.g.,][]{hansen2008effect, ajdacic2007seasonal, nori2022association}. This work builds on the small and recent body of economics literature that examines the causal impact of temperature on mental health \citep{burke2018higher, carleton2017crop, mullins2019temperature, obradovich2018empirical}. To the best of my knowledge, this the first work to use helpline data to study the link between temperature and mental health.

\noindent\\
A second major contribution of this paper is to provide insights into some of the primary mechanisms underlying the negative effects of temperature extremes on mental health. The channels through which mental health might be impaired by thermal stressors are complex and not yet well understood. There are a number of mechanisms that have been proposed so far. Potential direct mechanisms include the disruption of thermoregulatory function and alterations in central neurophysiological signaling \citep{lohmus2018possible}. Short-term exposure to unusual temperature levels may also indirectly compromise mental health through worsening physical health \citep{deschenes2014temperature}, temperature-related sleep disturbances \citep{obradovich2017nighttime, minor2022rising}, temperature-intolerance of antipsychotic medications \citep{cusack2011heatwaves}, decreasing cognitive functions \citep{graff2018temperature}, anxiety and stress due to anticipation of climate change (i.e., climate anxiety) \citep{clayton2020climate}, destructive behavior \citep{hsiang2013quantifying}, or mood alterations \citep{baylis2020temperature}, all of which are strongly correlated with mental health \citep{prince2007no, fernandez2013insomnia, thompson2011cross}. This is one of the first papers identifying the causal pathways behind this relationship, as few studies have empirically evaluated any of these claims \citep{massazza2022quantitative}.

\noindent\\
More broadly, this work joins two larger strands of literature. First, it adds to the rapidly growing climate economics literature which documents a range of negative consequences of human exposure to temperature extremes, including adverse effects on decision making \citep{heyes2019temperature}, sentiment \citep{baylis2020temperature}, violence and destructive behavior \citep{hsiang2013quantifying,almaas2019destructive, mukherjee2021causal}, cognitive attainment \citep{graff2018temperature, park2020hot}, physical performance \citep{sexton2021heat}, pregnancy outcomes \citep{deschenes2009climate, fishman2019long}, and more general health outcomes, such as mortality \citep{deschenes2011climate, barreca2016adapting} and morbidity \citep{white2017dynamic, karlsson2018population}. Second, this work contributes to the broad economics literature identifying causal determinants of mental health, including income \citep{gardner2007money}, indebtedness \citep{gathergood2012debt}, residential environment \citep{katz2001moving}, career and family choices \citep{bertrand2013career}, school environment \citep{butikofer2020school, kiessling2022long}, police violence \citep{ang2021effects}, prenatal stress \citep{persson2018family}, paid maternity leave \citep{butikofer2021impact}, early life stressors \citep{adhvaryu2019early}, social media usage \citep{braghieri2021social}, and air quality \citep{zhang2017happiness,chen2018air}.

\noindent\\
The remainder of this paper is organized as follows. Section 2 introduces the data. Section 3 presents the empirical approach. Section 4 reports and discusses the results. Section 5 concludes.

\section{Data and Summary Statistics} \label{sec:data}
The analysis relies primarily on three data sources. The first data source is high-resolution administrative data from a major European nonprofit telephone counseling service. The other two consist of the universe of measurements from German weather and air quality stations, which I use to interpolate daily environmental conditions at each site.

\noindent\\
\textbf{Helpline Data} - Individual contact-level data are provided by \textit{TelefonSeelsorge}, Germany's largest general telephone counseling service with more than 100 counseling centers, 300 employees, and nearly 7,700 volunteer counselors. The crisis service, founded in 1956, is well known\footnote{Panel A of Figure \ref{fig:Telefonseelsorge_google} in the Appendix shows the result of a representative online search for mental health concerns. Information about telephone counseling services is prominently placed on top of the landing page so that people seeking help for mental health problems can easily find out about the possible treatment option of telephone counseling. Panel B shows an example of an information poster.} and can be reached around the clock at one of two nationwide toll-free numbers, both from German landlines and mobile phones. Callers to the telephone counseling service are directed to the nearest crisis center, based on their location. The catchment area of each counseling center is defined by telephone area codes, with each crisis center serving surrounding phone codes in its vicinity. Figure \ref{fig:tel_munster} in the Appendix shows the extent of the catchment area of a representative counseling center operated by the helpline service.\footnote{Telephone area codes do not follow administrative boundaries and can sometimes even cross state lines.} Toward the end of 2018, the telephone counseling service introduced a new comprehensive data collection system, in which more than half of the crisis centers participate.\footnote{Each counseling center is operated independently, which is why at the time this study was conducted some sites in East Germany were using their own recording systems.} The data I receive was collected from November 2018 through the end of 2020. To obtain the final data set, I apply some sample selection criteria. First, because telephone counseling centers experienced a large increase in contact volume due to the COVID-19 pandemic \citep{brulhart2021distress}, I limit the observation period to before March 2020 to avoid bias.\footnote{The first containment measures in Germany were introduced on March 8, 2020.} Second, apart from telephone counseling, \textit{TelefonSeelsorge} offers counseling services via mail and chat. Since these contacts are routed nationwide, I focus on telephone counseling service only. 6.3 and 4.4 percent of all contacts recorded during the observation period were mail and chat contacts, respectively. 
\begin{figure}[t!]
\captionsetup{justification=centering,margin=1cm}
\centering
\caption{Map of helpline counseling centers in Germany}
\label{fig:summary_centers}
\includegraphics[width=.5\textwidth]{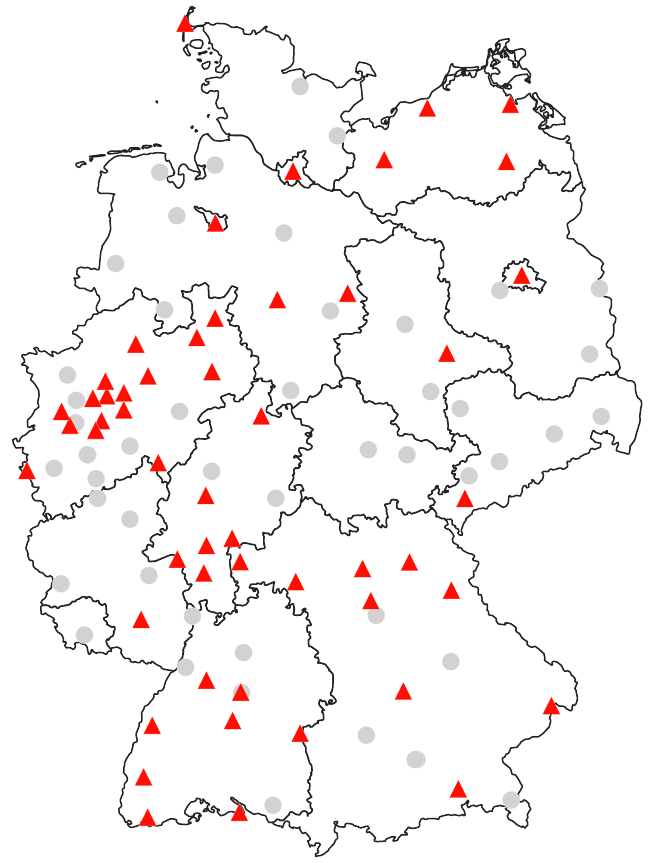}
\begin{minipage}{.8\textwidth} 
{\scriptsize{Notes: This figure shows the geographical distribution of \textit{TelefonSeelsorge}'s counseling centers in Germany. In total there are 105 counseling centers. Sampled locations are represented by a red triangle, non-sampled locations are represented by a grey dot.\par}}
\end{minipage}
\end{figure}
Third, as some sites only began using the main data collection system during the observation period, I exclude all locations that did not provide data on calls received on at least 100 days.\footnote{To be included in the sample, counseling centers must have begun using the main data system consecutively from November 11, 2019 onward. However, as shown in Table \ref{tab:summary_centers}, the majority of counseling centers already started collecting data in early 2019. The last site covered began collecting data in September 2019.} Based on this exclusion criteria, I exclude 20 counseling centers from the analysis that recorded a total of 27,175 calls, representing less than five percent of the total call volume before March 2020. Figure \ref{fig:summary_centers} shows the geographical distribution of all counseling centers in Germany. The sampled sites are distributed throughout the country, with most of the crisis centers studied located in West and South Germany. Individual call-level data include the purpose and duration of the call, sociodemographic information about the caller (i.e., gender, age group, employment status, and living situation)\footnote{\textit{TelefonSeelsorge} offers anonymous telephone counseling, which is why the information about the callers is based on the counselor's assessment. Some information is known from the contact (e.g., if the caller explicitly states her age), others are assumed from the context.}, topics of the conversation (i.e., physical and psychological well-being, violence, social well-being, livelihood, and other)\footnote{Operators categorize calls by the topics they discuss with callers, allowing them to select up to three topics from a list of 34 in total. To avoid a too extensive disaggregation, I aggregate the topics in six main groups. Table \ref{tab:calltopic} in the Appendix shows the classification of the call topics.}, and the counseling center where the call was answered. Finally, based on the recorded purpose of the call, I exclude all silent calls, prank calls, and sexually motivated calls which leads me to drop about one-fifth of the answered calls because they do not constitute help-seeking behavior for mental health problems.

\noindent\\
After applying the above exclusion criteria, I am left with 485,274 calls, that were answered at 55 counseling centers across Germany between November 3, 2018 and February 29, 2020. To obtain the number of calls answered at each location I collapse the individual calls to the daily level resulting in 21,138 counseling-center-day observations. 
Tables \ref{tab:summary_centers} and \ref{tab:summary_calls} in the Appendix provide a more detailed overview of the sampled crisis centers and the individual calls that were received by \textit{TelefonSeelsorge}. Figure \ref{fig:call_hist} and Figure \ref{fig:dur_hist} in the Appendix show the distribution of the daily number of calls answered and average helpline call duration, respectively. A crisis center answers an average of 22.96 calls per day, that last on average 1457.71 seconds (24 minutes, 18 seconds). The largest crisis center receives an average of 35.94 calls per day, while the smallest facility receives an average of 8.41 calls.

\begin{figure}[t!]
\centering
\captionsetup{justification=centering,margin=1cm}
\caption{Temperature distribution}
\label{fig:summary_temp}
\includegraphics[width=.8\textwidth]{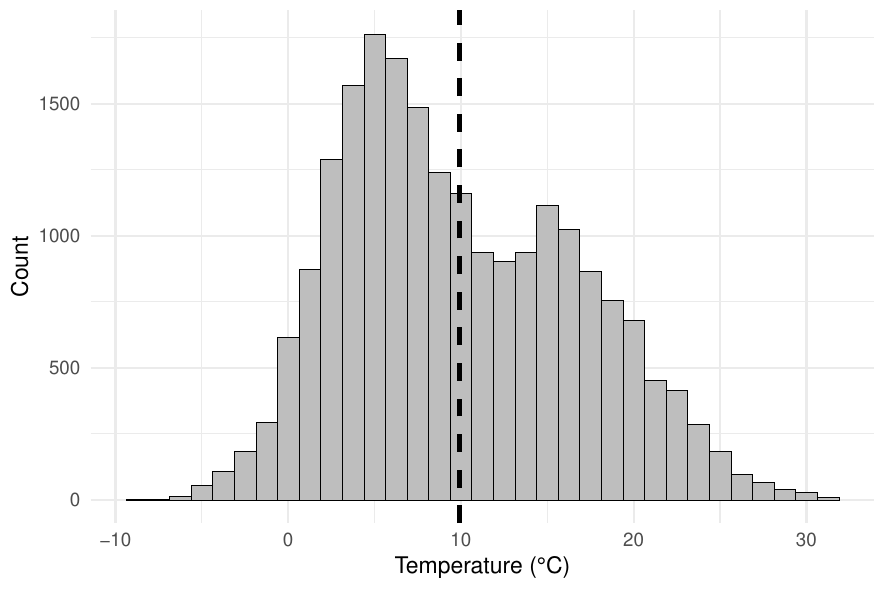}
\begin{minipage}{.8\textwidth} 
{\scriptsize{Notes: This figure illustrates the distribution of all daily mean temperatures at any given counseling-center-day in the study sample. The horizontal dashed line represents the sample mean (9.93$^{\circ}$C).\par}}
\end{minipage}
\end{figure}

\noindent\\
\textbf{Weather Data} - Station-level weather measurements are drawn from the Climate Data Center, which is a climate database operated by \textit{Deutscher Wetterdienst (DWD)}, the German Meteorological Service. I extract daily information on temperature, precipitation, sunshine duration, surface wind speed, and relative humidity.\footnote{In practice, I omit weather stations above 1,500 m since they are unlikely to represent the weather experienced by most of the population.} Since the telephone counseling service does not have information about the exact location of the caller, I take advantage of the fact that callers are routed to the nearest counseling center and use the weather conditions of the site that answered the call. To aggregate daily station-level weather realizations to the counseling-center-level, I take an inverse-distance weighted average of all observations provided by weather stations that are located within a predefined range to each counseling centers' location\footnote{\textit{TelefonSeelsorge} only discloses the zip code of the area where the site is located. Therefore, I use the centroid of the respective zip code area to calculate the distance to the weather stations.}, so that measured values closest to the counseling center have more influence on the predicted value than those farther away.\footnote{In total there are 1,924 precipitation monitoring stations, 502 temperature monitoring stations, 505 humidity monitoring stations, 287 surface wind speed monitoring stations, and 304 monitoring stations that provide information on sunshine duration. The choice of range is informed by the density of the respective monitoring network. I choose 15 km for precipitation, 25 km for temperature and humidity, and 35 km for wind speed and sunshine duration. An average counseling center is surrounded by 3.6 precipitation, 2.6 temperature, 2.5 humidity, 2.5 surface wind speed, as well as 2.9 sunshine duration stations in the specified radii.} Figure \ref{fig:summary_temp} illustrates the temperature distribution. The observation period provides a large temperature range to study the impact of temperature on mental health concerns, with the minimum average daily temperature of -8.3$^{\circ}$C recorded on January 23, 2019 in Auerbach (Saxonia) and the maximum daily average of 31.6$^{\circ}$C recorded on July 25, 2019 in Essen (North Rhine-Westphalia). Panel A - D of Figure \ref{fig:summary_envr} in the Appendix provide an overview of the distribution of other contemporaneous weather conditions realized within the study sample. 

\noindent\\
\textbf{Air Pollution Data} - Daily station-level concentration measurements of fine particulate matter with a diameter of less than 2.5 micrometers (PM$_{2.5}$) are obtained from the \textit{Umweltbundesamt (UBA)}, the German Federal Environmental Agency. I proceed in the same way as with the weather data and assign air quality information to each crisis center by taking an inverse-distance weighted average of observations of surrounding monitoring stations.\footnote{There is a total of 203 PM$_{2.5}$ monitoring stations that I use for the interpolation of counseling-center-level observations. As the monitoring network for PM$_{2.5}$ is not as dense as the weather monitoring network and is mainly located in urban areas, I have to choose a radius of 85 km so that all crisis centers are assigned at least one monitoring station. However, most of the counseling centers have a measurement station close to them, which has a larger impact on the predicted value due to the inverse distance weighting. The median distance to the nearest station is 7.9 km.} Panel E of Figure \ref{fig:summary_envr} in the Appendix shows the distribution of measured air pollution over the course of the study period.

\section{Empirical Approach}
To estimate the causal impact of ambient temperature on mental health I use a panel fixed effects model, as is standard in the climate economics literature \citep{dell2014we, hsiang2016climate}. I consider two different margins along which the use of telephone counseling service might be affected by exposure to abnormal temperature levels. In a first step I analyze if the incidence of helpline calls changes with temperature (i.e, the extensive margin). The model reads as follows:
\begin{equation} \label{eq:3}
y_{ct} =  f(\textrm{TEMP}_{ct}) +  \gamma W_{ct} + \alpha_{c} + \lambda_{t}  + \epsilon_{ct},
\end{equation}
where the dependent variable, $y_{ct}$, is the logarithm of the number of helpline calls answered in counseling center $c$ on day $t$. To avoid specifications of the relationship between average daily temperature levels and helpline call volumes, most of the time temperature enters the regression with a flexible functional form. More specifically, I let $f(\textrm{TEMP}_{ct}) = \sum_{\substack{j=1 \\ j\neq 4}}^{7} \beta^{j} \textrm{TEMP}^{j}_{ct}$ by dividing the temperature spectrum in seven five-degree intervals, ranging from less than 0$^{\circ}$C to more than 25$^{\circ}$C. The main variables of interest, $\textrm{TEMP}_{ct}^{j}$, indicate whether the average temperature on a given day at a given counseling center falls within the $j$th of the above mentioned temperature bands. I omit the temperature bin which is equal to unity if the average daily temperature is between 10$^{\circ}$C and 15$^{\circ}$C (i.e., $j \neq$ 4). As relative differences between conditional means are preserved, the choice of the omitted temperature range does not change the shape of the estimated response function. The coefficients $\beta^{j}$ show the relative change in helpline call volumes on a day in one of the temperature bins relative to a day in the omitted  category. 

\noindent\\
To account for the possibility that other weather dimensions besides temperature affect mental health, I include a vector of control variables, $W_{ct}$, that contains additional contemporaneous meteorological conditions, such as the sum of daily precipitation, daily sunshine duration, average daily wind speed, and average daily relative humidity. As mental health is potentially impaired by air quality \citep{zhang2017happiness} the vector of control variables also includes daily average concentrations of a key air pollutant (i.e., PM$_{2.5}$). $\alpha_{c}$ is a vector of location-specific fixed effects to capture permanent unobserved factors at the counseling center level, such as differences in recognition of mental health problems in the population served. $\lambda_{t}$ are day-of-week (including public holidays) and year-month fixed effects to control for trends within weeks, on holidays, and systematic seasonality. Assuming that the temperature-variation, conditional on the covariates and fixed effects, is exogenous to the unobserved determinants of the number of calls to counseling centers, the $\beta_{j}$ coefficients can be interpreted as the causal impact of temperature on mental health concerns as reflected in the demand for telephone counseling services. To allow for correlation of error terms standard errors are two-way clustered at the counseling center and year-month level. I further stratify the sample by demographic information of the caller and main topics of the conversation and re-estimate the regression for each group separately.\footnote{Since there are days when no calls are answered from a particular subgroup or on a particular topic, I sometimes use a log(helpline calls + 1) transformation. For the full sample there are no zero valued observations. Results obtained using alternative transformations of the dependent variable (e.g., inverse hyperbolic since (IHS)) and Poisson regression lead to the same conclusion  (Table \ref{tab:results_robust}).}

\noindent\\
As an additional empirical exercise, I investigate whether temperature affects the intensity of use of counseling services (i.e., the intensive margin). To do so, I use the logarithm of the duration of call $i$ answered at counseling center $c$ at time $t$, $z_{ict}$, as the dependent variable. I also include a vector of sociodemographic covariates, $X_{ict}$, to control for a set of individual-level caller characteristics that might correlate with the length of the counseling, including gender, age cohort, employment status and living situation. In this way, I can control for possible temperature-related changes in caller composition. To additionally control for trends within days, I include a full set of hour-of-day fixed effects. The rest of the model is similar to the one described in equation (\ref{eq:3}). Standard errors are clustered at the counseling-center-day level. In particular, I estimate the following regression equation using the full sample of individual call-level data:

\begin{equation} \label{eq:4}
z_{ict} =  f(\textrm{TEMP}_{ct}) +  \gamma W_{ct} + \delta X_{ict} + \alpha_{c} + \lambda_{t}  + \epsilon_{ict}.
\end{equation}

\noindent\\
There are some potential concerns related to the empirical strategy that I will discuss in more detail below. First, \textit{TelefonSeelsorge} does not have information on the exact location of the caller, so I leverage the fact that the caller is routed to the nearest counseling center and interpolate the weather data at that location. This introduces a potential measurement error in the explanatory variables, as the daily average temperature at a given counseling center is unlikely to be exactly the same as the temperature at the caller's site. However, since the telephone counseling service operates a fairly dense network of counseling centers\footnote{By comparison, while the largest telephone helpline in the United States (i.e., \textit{National Suicide Prevention Lifeline}) operates about 180 crisis centers across the country, \textit{TelefonSeelsorge} has a network of about 100 crisis centers. Yet Germany is approximately 28 times smaller in size than the United States.}, the extent of the catchment areas is relatively small and daily average temperatures are likely to be comparable in this area since temperature is rather uniform in space. In addition, the choice of wide temperature ranges provides a potential safeguard against the introduction of significant measurement error, as this increases the likelihood that the temperature at the caller's location and at the counseling center are at least within the same temperature interval. 

\noindent\\
Second, the relatively short observation period could be a cause for concern. However, the data I receive are of high spatial and temporal resolution. In addition, the year 2019, from which most of the observations come, was marked by high regional weather variability and extreme temperatures.\footnote{The year 2019 was the second warmest year since records began in 1881. Also, for the first time, a maximum temperature of 40$^{\circ}$C was measured for three consecutive days \citep{DWD2020}.} The high resolution of the helpline data in combination with the large weather fluctuations during the observation period provide a great variation for the model to be identified. As a falsification test, I conduct several placebo exercises. I replace the interpolated daily ambient weather realizations and air pollution measurement for each counseling center with the interpolated observations of the same day for the most distant counseling center. For example, for the crisis center in Berlin (Berlin), the placebo temperature is taken from Wehr (Baden-Wurttemberg), which is about 670 km away. Then I replace the measurements for each location with the observations for the same location, but recorded one quarter (90 days) before that day. For example, for the crisis center in Berlin, the placebo temperature for June 1, 2019 is taken from March 3, 2019. Finally, I do both, which means that the placebo temperature for the crisis center in Berlin on June 1, 2019 is taken from the crisis center in Wehr on March 3, 2019.

\noindent\\
Third, there is potential measurement error in the dependent variable. This does not bias the estimates but does affect their precision. While all calls made from a landline can be easily routed to the nearest counseling center based on the area code, the location-based routing of calls made from a cell phone is more difficult and is based on the location of the cell tower. As a result a small fraction of these calls is distributed randomly across Germany.\footnote{In August 2016, 78.4 percent of calls were routed according to the caller's location \citep{Telefonseelsorge2017}. For comparison, unlike  \textit{TelefonSeelsorge}, the \textit{National Suicide Prevention Lifeline} routes calls based on area code only, as all telephone numbers, whether landline or cell phone, have an area code. Due to the high mobility of cell phones, the area code of a mobile number may not reflect the actual location of the caller.} Additionally, the individual counseling centers are organized into organizational units consisting of three to seven sites that are geographically close to each other. While priority is given to routing a call to the nearest crisis center, if the call cannot be answered due to lack of available capacity, it is routed to another counseling center in the organizational unit. However, I cannot directly test if this affects the results, as the data does not provide any information on whether a call was diverted or not.\footnote{Another option would be to aggregate the call data at the organizational unit level and interpolate the environmental factors for the geographic center of gravity of each unit. However, I observe only four complete organizational units out of a total of 23.}

\noindent\\
Fourth, counseling centers might self-select for data collection. This does not pose a threat to identification, however, if participation in data collection is unrelated to the climatic conditions at the site, an assumption which is likely to hold.

\noindent\\
Finally, data from counseling centers may be an inaccurate indicator of mental health service demand and significantly underestimate the real need. However, if actual demand is even higher than it is reflected in the number of calls to counseling centers, the coefficients presented in this study represent a conservative estimate of the true effect of temperature on helpline call volumes and mental health concerns.

\section{Results}
\subsection{Baseline Results}

\begin{table}[t!] \centering 
\captionsetup{justification=centering,margin=1cm}
\centering 
  \caption{Effect of average daily temperature on helpline call volume} 
  \label{tab:results_main} 
\scriptsize
\begin{tabularx}{\textwidth}{lYYYY}
\\[-1.8ex]\hline 
\hline \\[-1.8ex] 
& \multicolumn{4}{c}{\textit{Dependent variable:}} \\ 
\cline{2-5} 
\\[-1.8ex] & \multicolumn{4}{c}{helpline calls} \\ 
\\[-1.8ex] & (1) & (2) & (3) & (4)\\ 
\hline \\[-1.8ex] 
\underline{\textit{Panel A. Nonlinear}} &  &  & & \\ 
 $<$ 0$^{\circ}$C  & 0.051$^{***}$ & 0.053$^{***}$ & 0.054$^{***}$ & 0.054$^{***}$ \\ 
  & (0.013) & (0.014) & (0.014) & (0.015) \\ 

 0$^{\circ}$C - 5$^{\circ}$C & 0.032$^{***}$ & 0.035$^{***}$ & 0.033$^{***}$ & 0.036$^{***}$ \\ 
  & (0.011) & (0.011) & (0.011) & (0.011) \\ 

 5$^{\circ}$C - 10$^{\circ}$C& 0.024$^{**}$ & 0.026$^{**}$ & 0.024$^{**}$ & 0.026$^{***}$ \\ 
  & (0.010) & (0.010) & (0.010) & (0.010) \\ 

10$^{\circ}$C - 15$^{\circ}$C & Ref. & Ref. & Ref. & Ref. \\ 
 & & & & \\ 

 15$^{\circ}$C - 20$^{\circ}$C & 0.004 & $-$0.0004 & 0.003 & $-$0.001 \\ 
  & (0.008) & (0.009) & (0.008) & (0.009) \\ 

 20$^{\circ}$C - 25$^{\circ}$C& $-$0.001 & $-$0.004 & $-$0.001 & $-$0.004 \\ 
  & (0.013) & (0.013) & (0.013) & (0.013) \\  

 $>$ 25$^{\circ}$C & 0.034$^{***}$ & 0.033$^{***}$ & 0.035$^{***}$ & 0.034$^{***}$ \\ 
  & (0.009) & (0.007) & (0.009) & (0.007) \\ 
  & & & &  \\ 

Observations & 21,138 & 21,138 & 21,138 & 21,138 \\ 
Adjusted R$^{2}$   & 0.455 & 0.454 & 0.455 & 0.454 \\ 
  & & & &   \\ 
\underline{\textit{Panel B. Linear}} &  &  & &   \\ 
 Temperature ($^{\circ}$C)& $-$0.001 & $-$0.001 & $-$0.001 & $-$0.002 \\ 
  & (0.001) & (0.001) & (0.001) & (0.001) \\
  &  &  & &   \\ 
Observations & 21,138 & 21,138 & 21,138 & 21,138  \\
Adjusted R$^{2}$ & 0.455 & 0.454 & 0.455 & 0.454 \\  
\hline \\[-1.8ex] 
Counseling-center fixed effects & Yes & Yes & Yes & Yes\\ 
Year-by-month fixed effects & Yes & Yes & Yes & Yes  \\ 
Environmental controls & Yes & Yes & Yes & Yes  \\
Day-of-week fixed effects & Yes & No & Yes & No\\ 
Holiday fixed effects & Yes & Yes & No & No \\ 
\hline 
\hline \\[-1.8ex] 
\multicolumn{5}{l}{Notes: The dependent variable is the natural logarithm of the daily number of answered helpline calls. The sample period} \\ 
\multicolumn{5}{l}{is November 3, 2018 to February 29, 2020. The standard errors in parentheses are two-way clustered at the counseling} \\
\multicolumn{5}{l}{center and year-month level. $^{*}$p$<$0.1; $^{**}$p$<$0.05; $^{***}$p$<$0.01.} \\
\end{tabularx}
\end{table}

\begin{figure}[t!]
\captionsetup{justification=centering,margin=1cm}
\centering
\caption{Estimated relationship between daily helpline call volume and average daily temperature}
\label{fig:estimates_temp}
\includegraphics[width=1\textwidth]{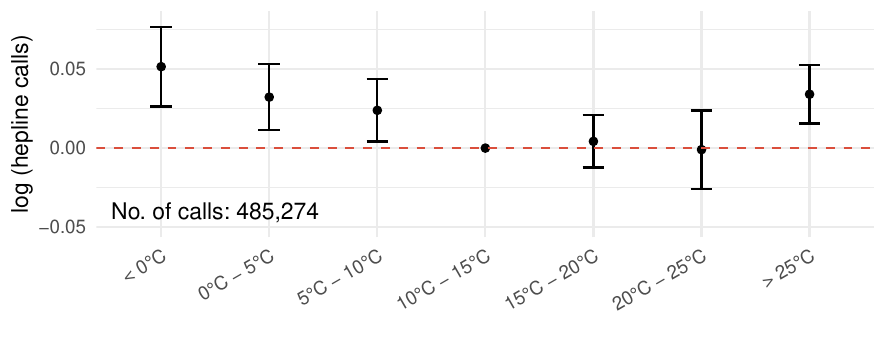}
\begin{minipage}{1\textwidth} 
{\scriptsize{Notes: This figure illustrates the response function between the natural logarithm of daily helpline call volumes and average daily temperatures (Table 1 Panel A column (1)). The response function is normalized with the 10$^{\circ}$C - 15$^{\circ}$C category set equal to zero. Each coefficient can be interpreted as the percentage change in the number of helpline calls on a day in bin $j$ relative to a day with an average temperature in the base category. Whiskers denote the obtained 95\% confidence intervals.\par}}
\end{minipage}
\end{figure}

\noindent
 In Table \ref{tab:results_main}, Panel A I present the results of different specifications of the model described in equation (\ref{eq:3}). I exclude specific fixed effects step by step to test for sensitivity of the results. Column (1) in Panel A of Table \ref{tab:results_main} presents the results for the main specification, where temperature enters the model with a flexible functional form. Although there are relatively few days in the upper and lower ranges of the daily temperature distribution, differences between call volumes on extreme-temperature days and days in the base category are statistically significant at conventional levels. The number of helpline calls answered in counseling centers is 5.1 percent higher on extremely cold days with an average temperature below 0$^{\circ}$C than on days with average temperatures between 10$^{\circ}$C and 15$^{\circ}$C. On the other end of the temperature distribution the demand for telephone counseling is 3.4 percent higher on extremely hot days with average temperatures above 25$^{\circ}$C, relative to days in the excluded temperature range. For comparison, on a day with an average temperature in the mid-temperature range a counseling center answers 22.8 calls, on average. Therefore, on extremely cold or hot days a crisis center answers an additional 1.2 or 0.8 calls, respectively. The coefficients increase toward the ends of the temperature distribution, indicating a U-shaped relationship between temperature and mental health problems. The number of calls answered on days with an average temperature in the middle range is not statistically different from the number of calls on a day in the excluded category. The magnitude of the point estimates is close to zero and statistically insignificant for days with average temperatures greater than 15$^{\circ}$C but less than 25$^{\circ}$C. The pattern still holds when day-of-week fixed effects and holiday fixed effects are further excluded in columns (2) - (4). As the focus of this paper lies on the effects of exposure to temperature on mental health, I do not report results for other weather dimensions or air pollution in the interest of space. In addition, other environmental factors are spatiotemporally highly heterogeneous compared to temperature, so the coefficients should be interpreted with caution as the exact location of the caller is not known. It is worth noting, however, that in the preferred specification (Table \ref{tab:results_main}, Panel A column (1)), I do not document statistically significant effects for weather realizations other than temperature. For ease of interpretation, Figure \ref{fig:estimates_temp} illustrates the response function between daily helpline call volumes and average daily ambient temperature. 

\noindent\\
Even though the results indicate a nonlinear relationship, I estimate a simpler version of the model where temperature enters the regression continuously. Results of that alternative continuous specification are provided in Panel B of Table \ref{tab:results_main}. The temperature coefficients are close to zero and insignificant across all specifications, an outcome that underscores the importance of considering more flexible responses of the outcome variable to temperature.

\noindent\\
The baseline estimates presented are consistent with previous work in that exposure to hot temperatures has adverse effects on mental health. However, the nonlinear shape of the estimated response function between the incidence of helpline calls and temperature is different from the previously studied relationship between temperature and more serious mental health outcomes, such as suicidality \citep{burke2018higher} and health care utilization \citep{mullins2019temperature}, as most studies employing traditional measures of mental health suggest a linear relationship. Instead, the U-shaped relationship mirrors that of temperature and general health outcomes \citep[e.g.,][]{deschenes2011climate}. Previous work looking at self-reported mental health also suggests nonlinear relationships between temperatures and mental health \citep{obradovich2018empirical}, with no clear evidence of temperature-related changes in mental health due to extreme cold. For comparison, \cite{obradovich2018empirical} report a 1.3 percent increase in the probability of self-reported mental health issues on days with an average maximum temperature above 30$^{\circ}$C relative to days with temperature maxima between 10$^{\circ}$C and 15$^{\circ}$C. \cite{burke2018higher} document a 0.7 percent increase in suicide rates for a 1$^{\circ}$C increase in monthly average temperatures.

\begin{figure}[t!]
\captionsetup{justification=centering,margin=1cm}
\centering
\caption{Estimated relationship between daily helpline call volume and average daily temperature by gender}
\label{fig:estimates_gend}
\includegraphics[width=1\textwidth]{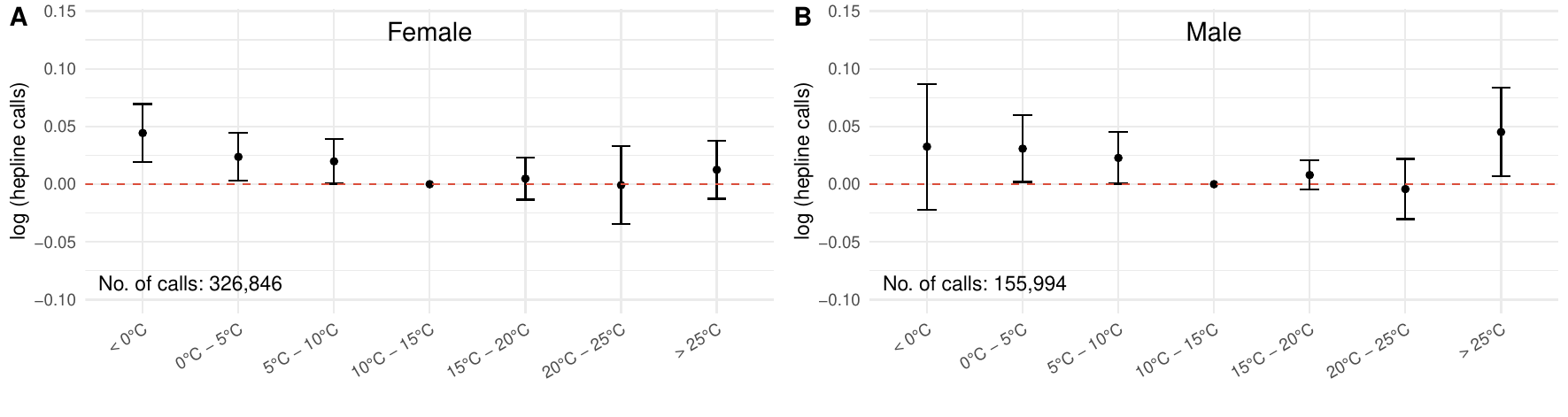}
\begin{minipage}{1\textwidth} 
{\scriptsize{Notes: This figure illustrates the response function between the natural logarithm of daily helpline call volumes and average daily temperatures by gender. Female (A); male (B), as estimated by applying the full model in equation (\ref{eq:3}) to the respective subsample. The response function is normalized with the 10$^{\circ}$C - 15$^{\circ}$C category set equal to zero. Each coefficient can be interpreted as the percentage change in the number of helpline calls on a day in bin $j$ relative to a day with an average temperature in the base category. Whiskers denote the obtained 95\% confidence intervals.\par}}
\end{minipage}
\end{figure}

\subsection{Heterogeneity}
\textbf{Heterogeneity by Gender} -
To examine the heterogeneous responses to temperature for mental health problems, I first stratify daily calls to helplines by gender. Figure \ref{fig:estimates_gend} plots the results when the sample is split by gender of the caller. While the estimates suggest an increase in call volumes on hot days for male callers relative to mild days by 4.5 percent, the number of calls to telephone counseling centers from female help-seekers on hot days does not differ from the number of calls on mid-temperature days. The magnitudes of the point estimates are closer to zero and statistically insignificant. However, on days with an ambient temperature in the lower end of the temperature spectrum the number of calls from women to helplines increases significantly relative to call volumes on days in the middle of the temperature range. Call volume for women is up by 4.4 percent on extremely cold days with average temperatures below 0$^{\circ}$C. Call volumes for male help-seekers also increase on colder days, but the magnitude of the effect for the outermost bin is smaller in size and the coefficient is imprecisely estimated, being insignificant on conventional levels.

\begin{figure}[t!]
\captionsetup{justification=centering,margin=1cm}
\centering
\caption{Estimated relationship between daily helpline call volume and average daily temperature by age}
\label{fig:estimates_age}
\includegraphics[width=1\textwidth]{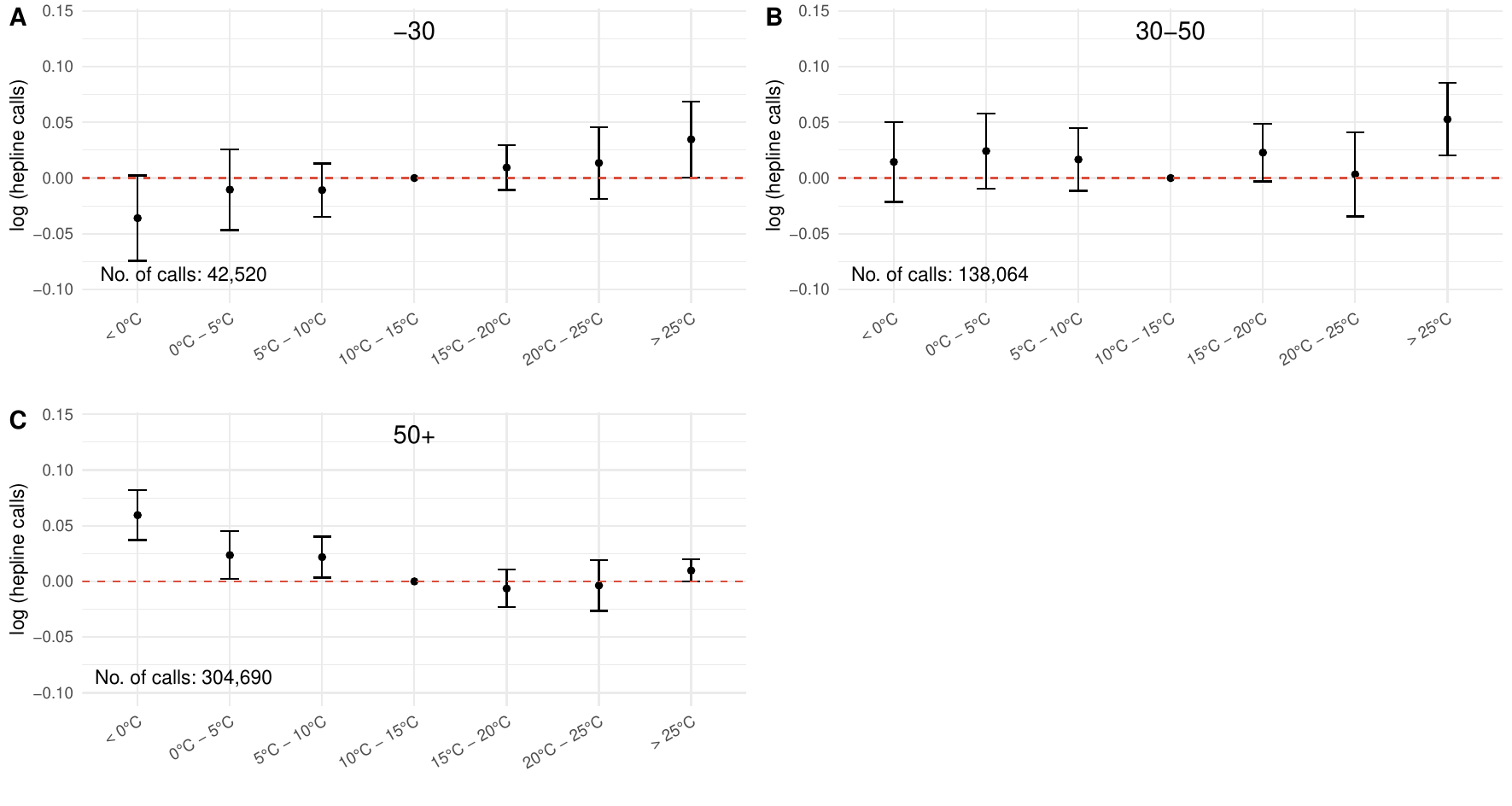}
\begin{minipage}{1\textwidth} 
{\scriptsize{Notes: This figure illustrates the response function between the natural logarithm of daily helpline call volumes and average daily temperatures by age. Younger than 30 years old (A); older than 30 but younger than 50 years old (B); older than 50 years old (C), as estimated by applying the full model in equation (\ref{eq:3}) to the respective subsample. The response function is normalized with the 10$^{\circ}$C - 15$^{\circ}$C category set equal to zero. Each coefficient can be interpreted as the percentage change in the number of helpline calls on a day in bin $j$ relative to a day with an average temperature in the base category. Whiskers denote the obtained 95\% confidence intervals.\par}}
\end{minipage}
\end{figure}

\noindent\\
\textbf{Heterogeneity by Age} -
Next, I study effect heterogeneity by dividing the sample in three age groups (i.e., 30-, 30-50, 50+). The results, shown in Figure \ref{fig:estimates_age}, suggest substantial age-related heterogeneity. The increase in call volume on cold days is caused mainly by people over 50 years of age (Panel C), who record an increase of up to 5.9 percent compared to days in the omitted category. The number of helpline calls from people under 50 years old on cold days does not differ systematically from the number of helpline calls on days with temperatures in the excluded range. However, I find that on extremely cold days, the number of calls from people below the age of 30 (Panel A) drops by 3.6 percent, albeit the coefficient is only weakly significant. At the other end of the temperature distribution, the overall increase in calls on warmer days is primarily attributable to younger individuals, with the impact on days with average temperatures above 25$^{\circ}$C being greatest for individuals between 30 and 50 years of age (Panel B). The results suggest a nearly linear relationship between temperature and mental health problems in people younger than 30.

\begin{figure}[b!]
\captionsetup{justification=centering,margin=1cm}
\centering
\caption{Estimated relationship between daily helpline call volume and average daily temperature by main topic}
\label{fig:estimates_top}
\includegraphics[width=1\textwidth]{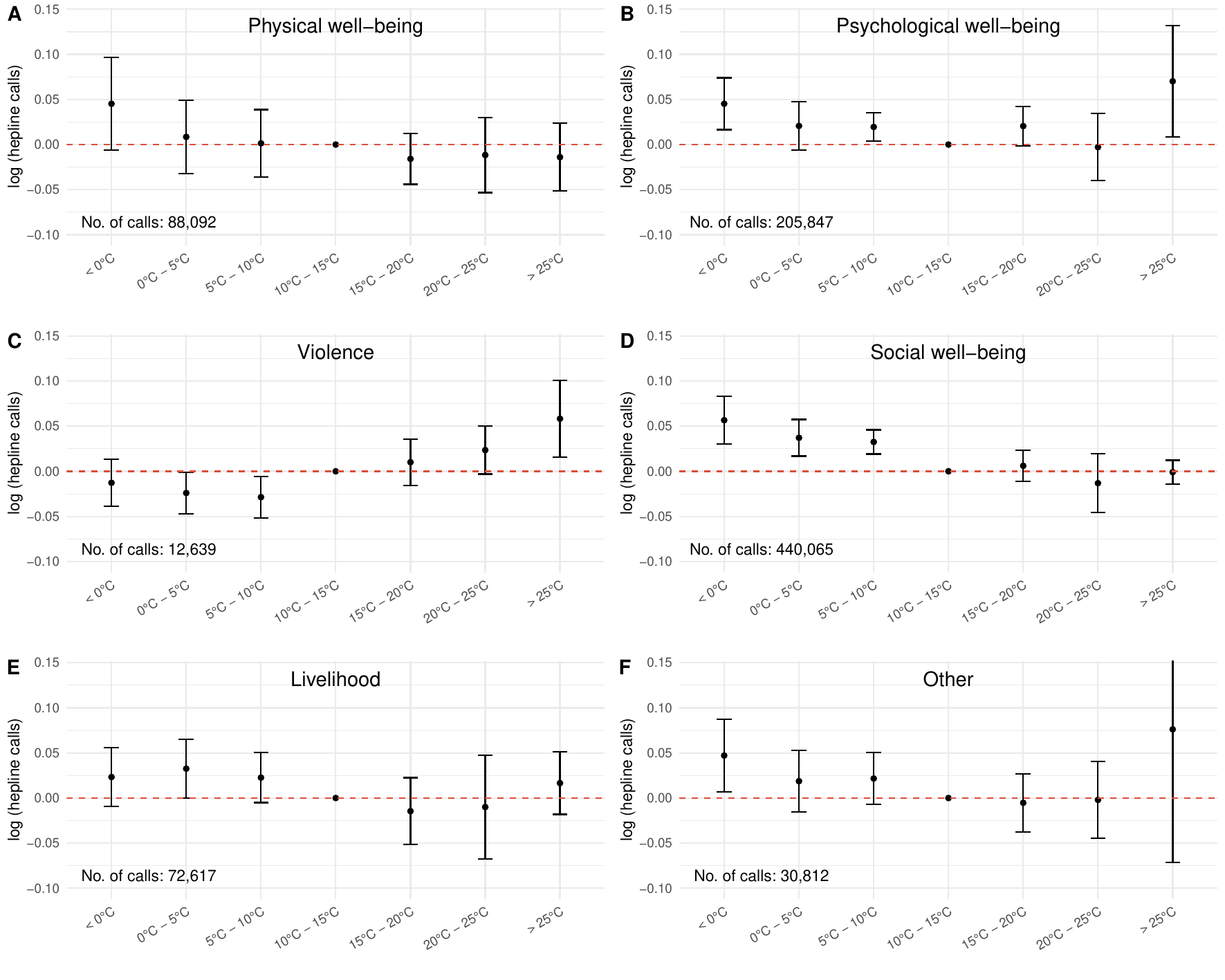}
\begin{minipage}{1\textwidth} 
{\scriptsize{Notes: This figure illustrates the response function between the natural logarithm of daily helpline call volumes and average daily temperatures by main topic of the conversation. Physical well-being (A); psychological well-being (B); violence (C); social well-being (D); livelihood (E); other (F), as estimated by applying the full model in equation (\ref{eq:3}) to the respective subsample. The response function is normalized with the 10$^{\circ}$C - 15$^{\circ}$C category set equal to zero. Each coefficient can be interpreted as the percentage change in the number of helpline calls on a day in bin $j$ relative to a day with an average temperature in the base category. Whiskers denote the obtained 95\% confidence intervals.\par}}
\end{minipage}
\end{figure}

\subsection{Mechanisms}

To analyze the underlying cause of concern of the consultation, I divide the sample according to the call`s main topic. In total there are 34 topics that operators can choose from. As mentioned above, I group the topics of the call into six main topics (Table \ref{tab:calltopic}). Results of the estimation are depicted in Panel A - F in Figure \ref{fig:estimates_top}. 

\noindent\\
\textbf{Psychological Well-being} - Panel B in Figure \ref{fig:estimates_top} plots the response function between the natural logarithm of daily helpline call volumes about psychological well-being (e.g., mood, self-harm, stress, self-image) and average daily temperatures. The results identify psychological well-being as an important factor for increasing demand for counseling services on extremely cold days. I find that the number of calls increases significantly on days at the low end of the temperature range by 4.5 percent. Similarly the number of calls increases at the high end of the temperature distribution by 7.0 percent, indicating that both extremely hot and cold temperatures indirectly undermine sound mental health through temperature-induced changes in psychological well-being. The nonlinear pattern, which emerges when I split the sample by topic and estimate the model for calls about psychological well-being only resembles evidence on temperature-related mood alterations. \cite{baylis2020temperature} analyzes the relationship between temperature and expressed mood using Twitter updates in the U.S. and finds a significant drop in expressed sentiment at both ends of the temperature spectrum. Thus, the estimates presented in the mechanism analysis support the existence of a causal psychological link between temperature exposure and mental health problems.

\noindent\\
\textbf{Violence} - I document that the number of calls for physical and sexual violence increases significantly by 5.8 percent on hot days compared to days with average temperatures (Panel C), with no increase on the lower end of the temperature range. If anything, the results suggest that the number of calls about violence is lower on colder days, however, the coefficient for the outermost bin is small in magnitude and not statistically significant on conventional levels. The results after estimating the model for physical and sexual violence calls only are consistent with the literature documenting a positive impact of heat exposure on interpersonal aggression, violence, and destructive behavior \citep{hsiang2013quantifying, almaas2019destructive, mukherjee2021causal}. For example, \cite{mukherjee2021causal} find that on days with high temperatures, the number of daily violent interactions among detainees increases by 20 percent, and the likelihood of any violent act  by up to 18 percent. \cite{almaas2019destructive} find that heat significantly affects the propensity for destructive actions in a laboratory setting. The negative effects in the lower temperature range could be explained by the fact that victims spend more time at home with their abusers and therefore cannot call the telephone counseling service \citep{leslie2020sheltering, anderberg2022quantifying}.

\noindent\\
\textbf{Social Well-being} -
The results suggest that on colder days, the number of calls about relationships (e.g., loneliness, partnership, family relationships) increases significantly by 5.7 percent compared to calls in average temperatures (Panel D). However, I find no evidence for an increase in calls about social relations on extremely hot days. The coefficients are close to zero and insignificant. Potentially the results can be explained by limited human mobility on colder days \citep{clarke2015impact}. People who live with others may spend more time at home during cold temperatures, which in turn could increase the likelihood of relationship problems. In addition, people who live alone may go out less and socialize less on cold days, which might increase feelings of loneliness and social isolation.  Thus, the estimates presented in the mechanism analysis suggest that cold temperatures might indirectly impair sound mental health by reducing social well-being.

\noindent\\
\textbf{Other Call Topics} -
The coefficients for the number of physical well-being and livelihood (e.g., work, poverty, finances) calls are not significantly different from zero (Panel A and E, respectively). However, the magnitudes of the coefficients and the shape of the response function for calls related to physical complaints provide suggestive evidence that the number of calls concerned with physical well-being increases on colder days compared to mid-temperature days. I estimate an increase of 4.5 percent for the outermost temperature range, a result that is only weakly statistically significant. In this particular setting, I do not find convincing evidence that exposure to hot temperatures indirectly compromise mental health through worsening physical health. In the short-term livelihood concerns do not seem to be affected by suboptimal temperatures, as the signs of the estimates are inconsistent and the magnitudes do not follow a consistent pattern. It intuitively seems plausible that livelihood concerns are not affected, as the effects presented in this study are short-run impacts of temperature on mental health. However, in the long-term, other temperature-induced effects, such as changes in economic conditions \citep{dell2012temperature} or migration \citep{deschenes2009extreme}, might have indirect impact on livelihood concerns.

\noindent\\
\textbf{Limitations} - As noted by \cite{liu2021helpline}, the analysis of call topics has some potential shortcomings as the stated topic of the call may not reflect the actual cause of concern, but rather may be a topic the caller wants to discuss with the counselor and the classification of the conversation topic may differ depending on the staff member making the assessment.\footnote{While this could be a major issue in cross-national studies, it is not as much of a concern in this work, as all counselors in telephone counseling go through the same training program before they begin their work in telephone counseling.}

\noindent\\
\textbf{Alternative Explanations} - There is a set of potential mechanisms that I cannot test directly in the data. Prior evidence suggests that warmer temperatures reduce sleep \citep{obradovich2017nighttime, minor2022rising} which in turn causally impacts human health \citep{jin2020sleep}. The relatively large effect size of minimum temperatures (Panel A Table \ref{tab:results_r2}), which are likely to be experienced during nighttime, provide suggestive evidence that temperatures impair mental health through sleep loss in this setting as well. Another alternative mechanism is given by temperature-related decreases in cognitive function. Interestingly, \cite{graff2018temperature} also find a nonlinear relationship between temperature and cognitive performance, that resembles the response function between temperature and mental health described in this work. Additionally, there are potential direct biological mechanisms through which temperature and mental health are linked, such as disruption of central processes involved in maintaining a stable body temperature \citep{lohmus2018possible}, that cannot be tested with the data at hand.

\begin{table}[b!] \centering 
  \caption{Effect of average daily temperature on helpline call volume, placebo}
 \label{tab:results_placebo} 
\scriptsize
\begin{tabularx}{\textwidth}{lYYY}
\\[-1.8ex]\hline 
\hline \\[-1.8ex] 
 & \multicolumn{3}{c}{\textit{Dependent variable:}} \\ 
\cline{2-4} 
\\[-1.8ex] & \multicolumn{3}{c}{helpline calls} \\ 
\\[-1.8ex] & furthest location & -90 days & furthest location \\ 
 &  &  & -90 days\\ 
\\[-1.8ex] & (1) & (2) & (3)\\ 
\hline \\[-1.8ex] 
 $<$ 0$^{\circ}$C & 0.004 & $-$0.020 & 0.015 \\ 
  & (0.018) & (0.020) & (0.022) \\   

0$^{\circ}$C - 5$^{\circ}$C& $-$0.006 & 0.003 & $-$0.003 \\ 
  & (0.011) & (0.014) & (0.012) \\ 
  
 5$^{\circ}$C - 10$^{\circ}$C & $-$0.005 & 0.016 & $-$0.002 \\ 
  & (0.012) & (0.013) & (0.011) \\ 

 10$^{\circ}$C - 15$^{\circ}$C& Ref. & Ref. & Ref.  \\ 
  & &&  \\ 

 15$^{\circ}$C - 20$^{\circ}$C & $-$0.006 & 0.002 & 0.001 \\ 
  & (0.018) & (0.009) & (0.016) \\

  20$^{\circ}$C - 25$^{\circ}$C& $-$0.023 & 0.011 & 0.006 \\ 
  & (0.026) & (0.012) & (0.015) \\ 
  
  $>$ 25$^{\circ}$C & 0.032 & $-$0.007 & 0.027 \\ 
  & (0.021) & (0.019) & (0.028) \\ 
& & & \\ 
Observations & 21,138 & 21,138 & 21,138 \\ 
Adjusted R$^{2}$ & 0.455 & 0.455 & 0.455 \\  
\hline \\[-1.8ex] 
Counseling-center fixed effects & Yes & Yes & Yes\\ 
Year-by-month fixed effects & Yes & Yes & Yes  \\ 
Environmental controls & Yes & Yes & Yes  \\ 
Day-of-week fixed effects & Yes & Yes & Yes \\ 
Holiday fixed effects & Yes & Yes & Yes \\ 
\hline 
\hline \\[-1.8ex] 
\multicolumn{4}{l}{Notes: The dependent variable is the natural logarithm of the daily number of answered helpline calls. The sample period} \\ 
\multicolumn{4}{l}{is November 3, 2018 to February 29, 2020. The standard errors in parentheses are two-way clustered at the counseling} \\
\multicolumn{4}{l}{center and year-month level. In column (1) placebo temperatures are taken from the counseling center that is the most} \\
\multicolumn{4}{l}{distant. In column (2) placebo temperatures are taken from 90 days earlier. In column(3) placebo temperatures are taken} \\
\multicolumn{4}{l}{from the most distant counseling center, but 90 days earlier. $^{*}$p$<$0.1; $^{**}$p$<$0.05;$^{***}$p$<$0.01.} \\
\end{tabularx} 
\end{table}

\subsection{Robustness and Validity}

I test the validity of the study design and the robustness of my results across multiple dimensions. My main results are consistent with changes in the benchmark specification. 

\noindent\\
\textbf{Placebo Tests} - Results of the falsification exercises, where I (i) replace daily environmental factors (i.e., temperature, precipitation, sunshine duration, wind speed, humidity, and air pollution) interpolated for each counseling center with the measurements of the same day but of the most distant site, (ii) replace daily environmental factors for each site with the observations recorded 90 days earlier, and (iii) replace daily environmental factors for each counseling center with the observations recorded 90 days earlier of the most distant site, are shown in Table \ref{tab:results_placebo}. Using a placebo test allows me to check for an association that should be present if the research design is flawed but not otherwise. In this placebo tests, the magnitude of the coefficients is closer to zero, the signs are inconsistent, and the estimates are insignificant in all specifications.\footnote{The choice of the 90 day backward shift is purely arbitrary. I repeat the exercises with periods of 60 and 120 days as well as with a forward shift. The results using these alternative placebo observations lead to the same conclusions, with mixed signs, estimates being closer to zero, and insignificant coefficients in all estimations. Results are available upon request.} I interpret the results of in-time and across-space placebos as supporting the causal interpretation of my estimates. In addition, I randomly permute indicators for extreme temperature days (i.e., $<$0$^{\circ}$C and $>$25$^{\circ}$C) across counseling-center-day observations 10,000 times, re-estimate the baseline model for each of these 10,000 generated dataset, and then compare the results to the observed estimate. Results of this simulation exercise are shown in Figure \ref{fig:summary_simul} in the Appendix. The simulation-based distribution of regression coefficients is centered at zero while the observed estimates are far to the right of this null distribution, providing further evidence that suggests there is a significant relationship between extreme temperatures and the number of calls to counseling centers.

\noindent\\
\textbf{Alternative Specifications and Clustering} -  I re-estimate the outcomes in column (1) of Table \ref{tab:results_main} using different transformations of the dependent variable. I estimate alternative specifications using the raw count of the number of helpline calls, the inverse hyperbolic sine (IHS) transformation of the number of helpline calls, and a Poisson specification and compare these to the baseline estimates. Results are shown in Table \ref{tab:results_robust} in the Appendix. The impact of extreme temperatures on the number of helpline calls remains highly significant and exhibits the same pattern of increasing intensity toward the end of the temperature range, suggesting that my choice of estimator does not drive the results.\footnote{The same applies to the models where I split the sample by gender, age, and topic. The results using alternative specifications mirror the results presented in main part of the paper.} In Table \ref{tab:results_robusterrors}, I explore alternative ways of clustering standard errors and compare these to the results when clustering at the counseling-center and year-month level. I cluster standard errors at larger areas to allow for spatial correlation of standard errors between counseling centers. Results remain robust when clustering at the federal state and year-month level and applying wild cluster bootstrapping to deal with the smaller number of resulting groups.\footnote{See Figure \ref{fig:summary_centers} for a visualisation of the spatial level of grouping. The black lines in the map show the borders of the German federal states.}

\noindent\\
\textbf{Environmental Conditions} - I provide evidence that the results are robust to the exclusion of simultaneous environmental conditions. I exclude one of the daily weather and pollution measurements at a time to check the sensitivity of the temperature coefficients to the individual contemporaneous environmental factors. Results are shown in Table \ref{tab:results_renv} in the Appendix. Estimated effects remain consistent with the baseline results.  

\noindent\\
\textbf{Temperature Exposure} - So far, I have relied on average ambient temperatures in order to remain agnostic as to when individuals are exposed to thermal stressors during the day. As additional robustness checks, I define exposure using daily maximum and minimum temperatures. As with average temperatures, the results indicate a pronounced increase in the number of helpline calls on days near the lower and upper ends of the respective temperature distribution. The results are shown in Panels A and B of Table \ref{tab:results_r2} in the Appendix.

\subsection{Dynamic Effects} In this section, I examine the dynamic effects of temperature on mental health. Up until now, this analysis has focused on on-the-day temperature impacts and ignored potential build up effects. It is possible that the increase in call volume on days with extreme temperatures is due to exposure to several days of suboptimal temperature levels. To this end, I follow the approach of \cite{somanathan2021impact}. I construct new variables indicating the number of days in the last three, five, or seven days that were extremely hot (i.e., $>$ 25$^{\circ}$C), and the number of days in the last three, five, or seven days that were extremely cold (i.e., $<$ 0$^{\circ}$C). To test whether past temperatures have an impact on today`s helpline call volume, I include the newly constructed variables in the baseline regression equation outlined in equation (\ref{eq:3}). The results of this empirical exercise are shown in Table \ref{tab:results_r3}. 

\begin{table}[b!] \centering 
\captionsetup{justification=centering,margin=1cm}
\centering 
  \caption{Dynamic effect of average daily temperature on helpline call volume} 
  \label{tab:results_r3} 
\scriptsize
\begin{tabularx}{\textwidth}{lYYY}
\\[-1.8ex]\hline 
\hline \\[-1.8ex] 
& \multicolumn{3}{c}{\textit{Dependent variable:}} \\ 
\cline {2-4}
\\[-1.8ex] & \multicolumn{3}{c}{helpline calls} \\ \\[-1.8ex] & 3 days & 5 days & 7 days \\ 
\\[-1.8ex]  & (1) & (2)& (3)\\ 
\hline \\[-1.8ex] 
$<$ 0$^{\circ}$C & 0.019 & 0.018 & 0.015 \\ 
  & (0.017) & (0.014) & (0.015) \\ 

 no. of previous days $<$ 0$^{\circ}$C & 0.011$^{***}$ & 0.009$^{***}$ & 0.006$^{**}$ \\ 
 & (0.004)  & (0.002)& (0.002) \\ 
  & & &\\  
 $>$ 25$^{\circ}$C & 0.024$^{***}$ & 0.024$^{*}$ & 0.027$^{**}$ \\ 
  & (0.009) & (0.013) & (0.011) \\  

 no. of previous days $>$ 25$^{\circ}$C &  0.011 & 0.008 & 0.002\\ 
&  (0.009) & (0.005) & (0.005) \\ 

  & & &\\  

Observations & 20,973 & 20,863 & 20,753 \\ 
Adjusted R$^{2}$ & 0.474 & 0.478 & 0.482 \\ 
\hline \\[-1.8ex] 
Counseling-center fixed effects &  Yes & Yes& Yes\\ 
Year-by-month fixed effects &  Yes  & Yes&Yes\\ 
Environmental controls &  Yes & Yes&Yes\\
Day-of-week fixed effects &  Yes &Yes&Yes\\ 
Holiday fixed effects &  Yes &Yes&Yes\\ 
\hline 
\hline \\[-1.8ex] 
\multicolumn{4}{l}{Notes: The dependent variable is the natural logarithm of the daily number of answered helpline calls. The sample period} \\ 
\multicolumn{4}{l}{is November 3, 2018 to February 29, 2020. The standard errors in parentheses are two-way clustered at the counseling} \\
\multicolumn{4}{l}{center and year-month level. No. of previous days refers to the count of days in the last (1) three, (2) five, and  (3) seven} \\
\multicolumn{4}{l}{days. $^{*}$p$<$0.1; $^{**}$p$<$0.05; $^{***}$p$<$0.01.} \\
\end{tabularx}
\end{table}

\noindent\\
I do not find convincing evidence of an impact of past exposure to hot temperature extremes on present mental health. Coefficients for the number of hot days in the previous three, five, and seven days are close to zero and statistically insignificant. Magnitude of point estimate for average on-the-day temperature is close to that of the estimates in the baseline specification in absolute value and highly significant. I interpret these findings as supporting evidence of the existence of an instant adverse effect of extremely hot temperatures on mental health. The results provide suggestive evidence of a cold build up. The coefficients for the counts of cold days in the past three, five, and seven days are significant across all specifications. For example, one additional cold day in the previous three days translates to a 1.1 percent increase in on-the-day call volume. The coefficient for the contemporaneous temperature is smaller in magnitude than in the base specification and imprecisely estimated.

\subsection{Other results} The results for the estimation of the model outlined in equation (\ref{eq:4}), where I employ the natural logarithm of call duration measured in seconds as the dependent variable are shown in Table \ref{tab:results_dur}. Results of the semiparametic specification (Panel A) suggest that calls on days with an average temperature in the lower end of the temperature distribution are 5.0 percent longer than calls on a day with an average temperature in the middle range. The magnitude of the estimates increases toward the lower end of the temperature distribution providing consistent evidence of a positive effect of low temperatures on call duration. The relationship between counseling length and warmer temperatures is less accurately estimated, although the negative sign of the coefficients suggests that hot temperatures lead to a reduction in the duration of telephone consultations. Again, I gradually exclude specific fixed effects to test the sensitivity of the results. Significance of estimates strongly depends on the model specification and vanishes when day-of-week, holiday and hour-of-day fixed effects are excluded. A look at the coefficient of the parsimonious model, where temperature enters the regression as a continuous measure, confirms the existence of an inverse linear relationship. In the preferred specification (Table \ref{tab:results_dur}, Panel B column (1)), an increase in temperature by 1$^{\circ}$C translates to a decrease in call duration by 0.3 percent, a finding that is statistically significant and robust across different specifications.

\begin{table}[t!] 
\captionsetup{justification=centering,margin=1cm}
\centering 
  \caption{Effect of average daily temperature on helpline call duration} 
  \label{tab:results_dur} 
\scriptsize
\begin{tabularx}{\textwidth}{lYYYYY}
\\[-1.8ex]\hline 
\hline \\[-1.8ex] 
 & \multicolumn{5}{c}{\textit{Dependent variable:}} \\ 
\cline{2-6} 
\\[-1.8ex] & \multicolumn{5}{c}{call duration} \\ 
\\[-1.8ex] & (1) & (2) & (3) & (4) & (5)\\ 
\hline \\[-1.8ex] 
\underline{\textit{Panel A. Nonlinear}} &  &  & &  & \\ 
 $<$ 0$^{\circ}$C  & 0.050$^{***}$ & 0.045$^{***}$ & 0.045$^{***}$ & 0.051$^{***}$ & 0.046$^{***}$ \\ 
  & (0.016) & (0.017) & (0.017) & (0.016) & (0.017) \\ 

 0$^{\circ}$C - 5$^{\circ}$C& 0.017$^{*}$ & 0.013 & 0.012 & 0.018$^{*}$ & 0.013 \\ 
  & (0.010) & (0.010) & (0.010) & (0.010) & (0.011) \\

 5$^{\circ}$C - 10$^{\circ}$C & 0.023$^{***}$ & 0.021$^{***}$ & 0.021$^{***}$ & 0.024$^{***}$ & 0.022$^{***}$ \\ 
  & (0.008) & (0.008) & (0.008) & (0.008) & (0.008) \\   
  
  10$^{\circ}$C - 15$^{\circ}$C & Ref. & Ref. & Ref. & Ref. & Ref.\\ 
 & & & & &\\ 

15$^{\circ}$C - 20$^{\circ}$C   & $-$0.010 & $-$0.004 & $-$0.003 & $-$0.009 & $-$0.003 \\ 
  & (0.013) & (0.013) & (0.013) & (0.013) & (0.013) \\   

20$^{\circ}$C - 25$^{\circ}$C& $-$0.053$^{*}$ & $-$0.044 & $-$0.044 & $-$0.052$^{*}$ & $-$0.043 \\ 
  & (0.031) & (0.030) & (0.030) & (0.031) & (0.030) \\    
 
 $>$ 25$^{\circ}$C & $-$0.012 & $-$0.014 & $-$0.014 & $-$0.013 & $-$0.014 \\ 
  & (0.022) & (0.023) & (0.023) & (0.022) & (0.023) \\ 
  &  &  & &  & \\ 
Observations & 485,274 & 485,274 & 485,274 & 485,274 & 485,274  \\ 
Adjusted R$^{2}$& 0.117 & 0.117 & 0.117 & 0.111 & 0.110 \\  
  &  &  & &  & \\ 
\underline{\textit{Panel B. Linear}} &  &  & &  & \\ 
 Temperature ($^{\circ}$C)& $-$0.003$^{***}$ & $-$0.003$^{***}$ & $-$0.003$^{***}$ & $-$0.003$^{***}$ & $-$0.003$^{***}$ \\ 
  & (0.001) & (0.001) & (0.001) & (0.001) & (0.001) \\  
  &  &  & &  & \\ 
  Observations & 485,274 & 485,274 & 485,274 & 485,274 & 485,274  \\ 
Adjusted R$^{2}$ & 0.117 & 0.117 & 0.117 & 0.110 & 0.110 \\ 
\hline \\[-1.8ex] 
Counseling-center fixed effects & Yes & Yes & Yes & Yes & Yes\\ 
Year-by-month fixed effects & Yes & Yes & Yes & Yes & Yes \\ 
Environmental controls& Yes & Yes & Yes & Yes & Yes \\ 
Individual controls & Yes & Yes & Yes & Yes & Yes \\
Day-of-week fixed effects & Yes & No & Yes & Yes & No\\ 
Holiday fixed effects & Yes & Yes & No & Yes & No\\ 
Hour-of-day fixed effects & Yes & Yes & Yes & No & No\\ 
\hline 
\hline \\[-1.8ex] 
\multicolumn{6}{l}{Notes: The dependent variable is the natural logarithm of the duration of answered helpline calls. The sample period is} \\ 
\multicolumn{6}{l}{November 3, 2018  to February 29, 2020. The standard errors in parentheses are two-way clustered at the counseling center} \\
\multicolumn{6}{l}{and date level.  $^{*}$p$<$0.1;$^{**}$p$<$0.05;$^{***}$p$<$0.01.} \\
\end{tabularx} 
\end{table} 

\noindent\\
The finding of shorter call duration on hot days might be explained by temperature-related changes in time preferences with increasing impatience as temperature rises, which, while not unambiguously, have been found to exist in previous research \citep{carias2021effects, almaas2019destructive}. However, I cannot rule out the possibility that the shorter duration of counseling is due to temperature-induced effects on telephone counselors, which may also be the cause of the shorter consultation duration. For example, \cite{somanathan2021impact} and \cite{lopalo2022temperature} indicate effects of hot temperatures on worker productivity. 

\section{Conclusion}
This paper investigates the causal link between temperature and mental health concerns. By using data on nearly half a million individual calls to Germany's largest telephone counseling service, I provide novel evidence of the adverse effects of exposure to extreme temperatures on mental health. Since mental health problems involve much more than just clinical diagnosis the use of helpline data makes an important contribution to our understanding of how environmental stressors might erode mental health. I document that the number of calls answered at counseling centers is significantly higher on extreme temperature days than on days with an average ambient daily temperature in the middle of the temperature distribution. I find significant age- and gender-related effect heterogeneity. Results suggest that increasing call volumes on days with a suboptimal temperature level are mainly due to an increase in calls related to social well-being, violence, and psychological well-being. Furthermore, I provide evidence of temperature-induced changes in the intensity of counseling use. Estimates show that the average length of telephone counseling decreases as temperature increases. The findings of this paper add to our understanding of how changes in climatic conditions may impact population mental health in the near future. The results highlight the large potential increase in social costs resulting from deteriorating mental health due to increased exposure to extreme temperatures and draw attention to another, previously overlooked, hidden cost of climate change.

\clearpage

\onehalfspacing

\bibliographystyle{apacite}
\bibliography{main.bib}

\begin{thebibliography}{}

\bibitem [\protect \citeauthoryear {%
Adhvaryu%
, Fenske%
\BCBL {}\ \BBA {} Nyshadham%
}{%
Adhvaryu%
\ \protect \BOthers {.}}{%
{\protect \APACyear {2019}}%
}]{%
adhvaryu2019early}
\APACinsertmetastar {%
adhvaryu2019early}%
\begin{APACrefauthors}%
Adhvaryu, A.%
, Fenske, J.%
\BCBL {}\ \BBA {} Nyshadham, A.%
\end{APACrefauthors}%
\unskip\
\newblock
\APACrefYearMonthDay{2019}{}{}.
\newblock
{\BBOQ}\APACrefatitle {Early life circumstance and adult mental health} {Early
  life circumstance and adult mental health}.{\BBCQ}
\newblock
\APACjournalVolNumPages{Journal of Political Economy}{127}{4}{1516--1549}.
\PrintBackRefs{\CurrentBib}

\bibitem [\protect \citeauthoryear {%
Ajdacic-Gross%
\ \protect \BOthers {.}}{%
Ajdacic-Gross%
\ \protect \BOthers {.}}{%
{\protect \APACyear {2007}}%
}]{%
ajdacic2007seasonal}
\APACinsertmetastar {%
ajdacic2007seasonal}%
\begin{APACrefauthors}%
Ajdacic-Gross, V.%
, Lauber, C.%
, Sansossio, R.%
, Bopp, M.%
, Eich, D.%
, Gostynski, M.%
\BDBL {}R{\"o}ssler, W.%
\end{APACrefauthors}%
\unskip\
\newblock
\APACrefYearMonthDay{2007}{}{}.
\newblock
{\BBOQ}\APACrefatitle {Seasonal associations between weather conditions and
  suicide—evidence against a classic hypothesis} {Seasonal associations
  between weather conditions and suicide—evidence against a classic
  hypothesis}.{\BBCQ}
\newblock
\APACjournalVolNumPages{American Journal of Epidemiology}{165}{5}{561--569}.
\PrintBackRefs{\CurrentBib}

\bibitem [\protect \citeauthoryear {%
Alm{\aa}s%
\ \protect \BOthers {.}}{%
Alm{\aa}s%
\ \protect \BOthers {.}}{%
{\protect \APACyear {2019}}%
}]{%
almaas2019destructive}
\APACinsertmetastar {%
almaas2019destructive}%
\begin{APACrefauthors}%
Alm{\aa}s, I.%
, Auffhammer, M.%
, Bold, T.%
, Bolliger, I.%
, Dembo, A.%
, Hsiang, S\BPBI M.%
\BDBL {}Pickmans, R.%
\end{APACrefauthors}%
\unskip\
\newblock
\APACrefYearMonthDay{2019}{}{}.
\newblock
{\BBOQ}\APACrefatitle {Destructive behavior, judgment, and economic
  decision-making under thermal stress} {Destructive behavior, judgment, and
  economic decision-making under thermal stress}.{\BBCQ}
\newblock
\APACjournalVolNumPages{NBER Working Paper No. 25785}{}{}{}.
\PrintBackRefs{\CurrentBib}

\bibitem [\protect \citeauthoryear {%
Anderberg%
, Rainer%
\BCBL {}\ \BBA {} Siuda%
}{%
Anderberg%
\ \protect \BOthers {.}}{%
{\protect \APACyear {2022}}%
}]{%
anderberg2022quantifying}
\APACinsertmetastar {%
anderberg2022quantifying}%
\begin{APACrefauthors}%
Anderberg, D.%
, Rainer, H.%
\BCBL {}\ \BBA {} Siuda, F.%
\end{APACrefauthors}%
\unskip\
\newblock
\APACrefYearMonthDay{2022}{}{}.
\newblock
{\BBOQ}\APACrefatitle {Quantifying domestic violence in times of crisis: An
  internet search activity-based measure for the COVID-19 pandemic}
  {Quantifying domestic violence in times of crisis: An internet search
  activity-based measure for the covid-19 pandemic}.{\BBCQ}
\newblock
\APACjournalVolNumPages{Journal of the Royal Statistical Society, Series A
  (Statistics in Society)}{}{}{}.
\PrintBackRefs{\CurrentBib}

\bibitem [\protect \citeauthoryear {%
Ang%
}{%
Ang%
}{%
{\protect \APACyear {2021}}%
}]{%
ang2021effects}
\APACinsertmetastar {%
ang2021effects}%
\begin{APACrefauthors}%
Ang, D.%
\end{APACrefauthors}%
\unskip\
\newblock
\APACrefYearMonthDay{2021}{}{}.
\newblock
{\BBOQ}\APACrefatitle {The effects of police violence on inner-city students}
  {The effects of police violence on inner-city students}.{\BBCQ}
\newblock
\APACjournalVolNumPages{The Quarterly Journal of Economics}{136}{1}{115--168}.
\PrintBackRefs{\CurrentBib}

\bibitem [\protect \citeauthoryear {%
Armbruster%
\ \BBA {} Klotzb{\"u}cher%
}{%
Armbruster%
\ \BBA {} Klotzb{\"u}cher%
}{%
{\protect \APACyear {2020}}%
}]{%
armbruster2020lost}
\APACinsertmetastar {%
armbruster2020lost}%
\begin{APACrefauthors}%
Armbruster, S.%
\BCBT {}\ \BBA {} Klotzb{\"u}cher, V.%
\end{APACrefauthors}%
\unskip\
\newblock
\APACrefYearMonthDay{2020}{}{}.
\newblock
{\BBOQ}\APACrefatitle {Lost in lockdown? {COVID-19}, social distancing, and
  mental health in Germany} {Lost in lockdown? {COVID-19}, social distancing,
  and mental health in germany}.{\BBCQ}
\newblock
\APACjournalVolNumPages{Covid Economics, Vetted and Real-Time
  Papers}{22}{}{117--153}.
\PrintBackRefs{\CurrentBib}

\bibitem [\protect \citeauthoryear {%
Barreca%
, Clay%
, Desch{\^e}nes%
, Greenstone%
\BCBL {}\ \BBA {} Shapiro%
}{%
Barreca%
\ \protect \BOthers {.}}{%
{\protect \APACyear {2016}}%
}]{%
barreca2016adapting}
\APACinsertmetastar {%
barreca2016adapting}%
\begin{APACrefauthors}%
Barreca, A.%
, Clay, K.%
, Desch{\^e}nes, O.%
, Greenstone, M.%
\BCBL {}\ \BBA {} Shapiro, J\BPBI S.%
\end{APACrefauthors}%
\unskip\
\newblock
\APACrefYearMonthDay{2016}{}{}.
\newblock
{\BBOQ}\APACrefatitle {Adapting to climate change: The remarkable decline in
  the US temperature-mortality relationship over the twentieth century}
  {Adapting to climate change: The remarkable decline in the us
  temperature-mortality relationship over the twentieth century}.{\BBCQ}
\newblock
\APACjournalVolNumPages{Journal of Political Economy}{124}{1}{105--159}.
\PrintBackRefs{\CurrentBib}

\bibitem [\protect \citeauthoryear {%
Baylis%
}{%
Baylis%
}{%
{\protect \APACyear {2020}}%
}]{%
baylis2020temperature}
\APACinsertmetastar {%
baylis2020temperature}%
\begin{APACrefauthors}%
Baylis, P.%
\end{APACrefauthors}%
\unskip\
\newblock
\APACrefYearMonthDay{2020}{}{}.
\newblock
{\BBOQ}\APACrefatitle {Temperature and temperament: Evidence from {Twitter}}
  {Temperature and temperament: Evidence from {Twitter}}.{\BBCQ}
\newblock
\APACjournalVolNumPages{Journal of Public Economics}{184}{}{104161}.
\PrintBackRefs{\CurrentBib}

\bibitem [\protect \citeauthoryear {%
Berry%
, Waite%
, Dear%
, Capon%
\BCBL {}\ \BBA {} Murray%
}{%
Berry%
\ \protect \BOthers {.}}{%
{\protect \APACyear {2018}}%
}]{%
berry2018case}
\APACinsertmetastar {%
berry2018case}%
\begin{APACrefauthors}%
Berry, H\BPBI L.%
, Waite, T\BPBI D.%
, Dear, K\BPBI B.%
, Capon, A\BPBI G.%
\BCBL {}\ \BBA {} Murray, V.%
\end{APACrefauthors}%
\unskip\
\newblock
\APACrefYearMonthDay{2018}{}{}.
\newblock
{\BBOQ}\APACrefatitle {The case for systems thinking about climate change and
  mental health} {The case for systems thinking about climate change and mental
  health}.{\BBCQ}
\newblock
\APACjournalVolNumPages{Nature Climate Change}{8}{4}{282--290}.
\PrintBackRefs{\CurrentBib}

\bibitem [\protect \citeauthoryear {%
Bertrand%
}{%
Bertrand%
}{%
{\protect \APACyear {2013}}%
}]{%
bertrand2013career}
\APACinsertmetastar {%
bertrand2013career}%
\begin{APACrefauthors}%
Bertrand, M.%
\end{APACrefauthors}%
\unskip\
\newblock
\APACrefYearMonthDay{2013}{}{}.
\newblock
{\BBOQ}\APACrefatitle {Career, family, and the well-being of college-educated
  women} {Career, family, and the well-being of college-educated women}.{\BBCQ}
\newblock
\APACjournalVolNumPages{American Economic Review}{103}{3}{244--50}.
\PrintBackRefs{\CurrentBib}

\bibitem [\protect \citeauthoryear {%
Bharadwaj%
, Pai%
\BCBL {}\ \BBA {} Suziedelyte%
}{%
Bharadwaj%
\ \protect \BOthers {.}}{%
{\protect \APACyear {2017}}%
}]{%
bharadwaj2017mental}
\APACinsertmetastar {%
bharadwaj2017mental}%
\begin{APACrefauthors}%
Bharadwaj, P.%
, Pai, M\BPBI M.%
\BCBL {}\ \BBA {} Suziedelyte, A.%
\end{APACrefauthors}%
\unskip\
\newblock
\APACrefYearMonthDay{2017}{}{}.
\newblock
{\BBOQ}\APACrefatitle {Mental health stigma} {Mental health stigma}.{\BBCQ}
\newblock
\APACjournalVolNumPages{Economics Letters}{159}{}{57--60}.
\PrintBackRefs{\CurrentBib}

\bibitem [\protect \citeauthoryear {%
Biasi%
, Dahl%
\BCBL {}\ \BBA {} Moser%
}{%
Biasi%
\ \protect \BOthers {.}}{%
{\protect \APACyear {2021}}%
}]{%
biasi2021career}
\APACinsertmetastar {%
biasi2021career}%
\begin{APACrefauthors}%
Biasi, B.%
, Dahl, M\BPBI S.%
\BCBL {}\ \BBA {} Moser, P.%
\end{APACrefauthors}%
\unskip\
\newblock
\APACrefYearMonthDay{2021}{}{}.
\newblock
{\BBOQ}\APACrefatitle {Career effects of mental health} {Career effects of
  mental health}.{\BBCQ}
\newblock
\APACjournalVolNumPages{NBER Working Paper No. 29031}{}{}{}.
\PrintBackRefs{\CurrentBib}

\bibitem [\protect \citeauthoryear {%
Bloom%
\ \protect \BOthers {.}}{%
Bloom%
\ \protect \BOthers {.}}{%
{\protect \APACyear {2011}}%
}]{%
WEF2011}
\APACinsertmetastar {%
WEF2011}%
\begin{APACrefauthors}%
Bloom, D.%
, Cafiero, E.%
, Jané-Llopis, E.%
, Abrahams-Gessel, S.%
, Bloom, L.%
, Fathima, A., S.~Feigl%
\BDBL {}Weinstein, C.%
\end{APACrefauthors}%
\unskip\
\newblock
\APACrefYear{2011}.
\newblock
\APACrefbtitle {The Global Economic Burden of Noncommunicable Diseases.} {The
  global economic burden of noncommunicable diseases.}
\newblock
\APACaddressPublisher{}{World Economic Forum}.
\PrintBackRefs{\CurrentBib}

\bibitem [\protect \citeauthoryear {%
Braghieri%
, Levy%
\BCBL {}\ \BBA {} Makarin%
}{%
Braghieri%
\ \protect \BOthers {.}}{%
{\protect \APACyear {2022}}%
}]{%
braghieri2021social}
\APACinsertmetastar {%
braghieri2021social}%
\begin{APACrefauthors}%
Braghieri, L.%
, Levy, R.%
\BCBL {}\ \BBA {} Makarin, A.%
\end{APACrefauthors}%
\unskip\
\newblock
\APACrefYearMonthDay{2022}{}{}.
\newblock
{\BBOQ}\APACrefatitle {Social Media and Mental Health} {Social media and mental
  health}.{\BBCQ}
\newblock
\APACjournalVolNumPages{CESifo Working Paper No. 9723}{}{}{}.
\PrintBackRefs{\CurrentBib}

\bibitem [\protect \citeauthoryear {%
Br{\"u}lhart%
, Klotzb{\"u}cher%
, Lalive%
\BCBL {}\ \BBA {} Reich%
}{%
Br{\"u}lhart%
\ \protect \BOthers {.}}{%
{\protect \APACyear {2021}}%
}]{%
brulhart2021distress}
\APACinsertmetastar {%
brulhart2021distress}%
\begin{APACrefauthors}%
Br{\"u}lhart, M.%
, Klotzb{\"u}cher, V.%
, Lalive, R.%
\BCBL {}\ \BBA {} Reich, S.%
\end{APACrefauthors}%
\unskip\
\newblock
\APACrefYearMonthDay{2021}{}{}.
\newblock
{\BBOQ}\APACrefatitle {Mental Health Concerns during {COVID-19} as Revealed
  through Helpline Calls} {Mental health concerns during {COVID-19} as revealed
  through helpline calls}.{\BBCQ}
\newblock
\APACjournalVolNumPages{Nature}{600}{}{1--6}.
\PrintBackRefs{\CurrentBib}

\bibitem [\protect \citeauthoryear {%
Burke%
\ \protect \BOthers {.}}{%
Burke%
\ \protect \BOthers {.}}{%
{\protect \APACyear {2018}}%
}]{%
burke2018higher}
\APACinsertmetastar {%
burke2018higher}%
\begin{APACrefauthors}%
Burke, M.%
, Gonz{\'a}lez, F.%
, Baylis, P.%
, Heft-Neal, S.%
, Baysan, C.%
, Basu, S.%
\BCBL {}\ \BBA {} Hsiang, S.%
\end{APACrefauthors}%
\unskip\
\newblock
\APACrefYearMonthDay{2018}{}{}.
\newblock
{\BBOQ}\APACrefatitle {Higher temperatures increase suicide rates in the
  {United States} and {Mexico}} {Higher temperatures increase suicide rates in
  the {United States} and {Mexico}}.{\BBCQ}
\newblock
\APACjournalVolNumPages{Nature climate change}{8}{8}{723--729}.
\PrintBackRefs{\CurrentBib}

\bibitem [\protect \citeauthoryear {%
B{\"u}tikofer%
, Cronin%
\BCBL {}\ \BBA {} Skira%
}{%
B{\"u}tikofer%
\ \protect \BOthers {.}}{%
{\protect \APACyear {2020}}%
}]{%
butikofer2020employment}
\APACinsertmetastar {%
butikofer2020employment}%
\begin{APACrefauthors}%
B{\"u}tikofer, A.%
, Cronin, C\BPBI J.%
\BCBL {}\ \BBA {} Skira, M\BPBI M.%
\end{APACrefauthors}%
\unskip\
\newblock
\APACrefYearMonthDay{2020}{}{}.
\newblock
{\BBOQ}\APACrefatitle {Employment effects of healthcare policy: Evidence from
  the 2007 FDA black box warning on antidepressants} {Employment effects of
  healthcare policy: Evidence from the 2007 fda black box warning on
  antidepressants}.{\BBCQ}
\newblock
\APACjournalVolNumPages{Journal of Health Economics}{73}{}{102348}.
\PrintBackRefs{\CurrentBib}

\bibitem [\protect \citeauthoryear {%
Butikofer%
, Ginja%
, Landaud%
\BCBL {}\ \BBA {} L{\o}ken%
}{%
Butikofer%
\ \protect \BOthers {.}}{%
{\protect \APACyear {2020}}%
}]{%
butikofer2020school}
\APACinsertmetastar {%
butikofer2020school}%
\begin{APACrefauthors}%
Butikofer, A.%
, Ginja, R.%
, Landaud, F.%
\BCBL {}\ \BBA {} L{\o}ken, K\BPBI V.%
\end{APACrefauthors}%
\unskip\
\newblock
\APACrefYearMonthDay{2020}{}{}.
\newblock
{\BBOQ}\APACrefatitle {School selectivity, peers, and mental health} {School
  selectivity, peers, and mental health}.{\BBCQ}
\newblock
\APACjournalVolNumPages{NHH Dept. of Economics Discussion Paper}{}{21}{}.
\PrintBackRefs{\CurrentBib}

\bibitem [\protect \citeauthoryear {%
B{\"u}tikofer%
, Riise%
\BCBL {}\ \BBA {} Skira%
}{%
B{\"u}tikofer%
\ \protect \BOthers {.}}{%
{\protect \APACyear {2021}}%
}]{%
butikofer2021impact}
\APACinsertmetastar {%
butikofer2021impact}%
\begin{APACrefauthors}%
B{\"u}tikofer, A.%
, Riise, J.%
\BCBL {}\ \BBA {} Skira, M\BPBI M.%
\end{APACrefauthors}%
\unskip\
\newblock
\APACrefYearMonthDay{2021}{}{}.
\newblock
{\BBOQ}\APACrefatitle {The impact of paid maternity leave on maternal health}
  {The impact of paid maternity leave on maternal health}.{\BBCQ}
\newblock
\APACjournalVolNumPages{American Economic Journal: Economic
  Policy}{13}{1}{67--105}.
\PrintBackRefs{\CurrentBib}

\bibitem [\protect \citeauthoryear {%
Carias%
, Johnston%
, Knott%
\BCBL {}\ \BBA {} Sweeney%
}{%
Carias%
\ \protect \BOthers {.}}{%
{\protect \APACyear {2021}}%
}]{%
carias2021effects}
\APACinsertmetastar {%
carias2021effects}%
\begin{APACrefauthors}%
Carias, M\BPBI E.%
, Johnston, D.%
, Knott, R.%
\BCBL {}\ \BBA {} Sweeney, R.%
\end{APACrefauthors}%
\unskip\
\newblock
\APACrefYearMonthDay{2021}{}{}.
\newblock
{\BBOQ}\APACrefatitle {The Effects of Temperature on Economic Preferences} {The
  effects of temperature on economic preferences}.{\BBCQ}
\newblock
\APACjournalVolNumPages{arXiv preprint arXiv:2110.05611}{}{}{}.
\PrintBackRefs{\CurrentBib}

\bibitem [\protect \citeauthoryear {%
Carleton%
}{%
Carleton%
}{%
{\protect \APACyear {2017}}%
}]{%
carleton2017crop}
\APACinsertmetastar {%
carleton2017crop}%
\begin{APACrefauthors}%
Carleton, T\BPBI A.%
\end{APACrefauthors}%
\unskip\
\newblock
\APACrefYearMonthDay{2017}{}{}.
\newblock
{\BBOQ}\APACrefatitle {Crop-damaging temperatures increase suicide rates in
  India} {Crop-damaging temperatures increase suicide rates in india}.{\BBCQ}
\newblock
\APACjournalVolNumPages{{Proceedings of the National Academy of
  Sciences}}{114}{33}{8746--8751}.
\PrintBackRefs{\CurrentBib}

\bibitem [\protect \citeauthoryear {%
Chen%
, Oliva%
\BCBL {}\ \BBA {} Zhang%
}{%
Chen%
\ \protect \BOthers {.}}{%
{\protect \APACyear {2018}}%
}]{%
chen2018air}
\APACinsertmetastar {%
chen2018air}%
\begin{APACrefauthors}%
Chen, S.%
, Oliva, P.%
\BCBL {}\ \BBA {} Zhang, P.%
\end{APACrefauthors}%
\unskip\
\newblock
\APACrefYearMonthDay{2018}{}{}.
\newblock
{\BBOQ}\APACrefatitle {Air pollution and mental health: Evidence from China}
  {Air pollution and mental health: Evidence from china}.{\BBCQ}
\newblock
\APACjournalVolNumPages{NBER Working Paper No. 24686}{}{}{}.
\PrintBackRefs{\CurrentBib}

\bibitem [\protect \citeauthoryear {%
Choi%
\ \protect \BOthers {.}}{%
Choi%
\ \protect \BOthers {.}}{%
{\protect \APACyear {2020}}%
}]{%
choi2020development}
\APACinsertmetastar {%
choi2020development}%
\begin{APACrefauthors}%
Choi, D.%
, Sumner, S\BPBI A.%
, Holland, K\BPBI M.%
, Draper, J.%
, Murphy, S.%
, Bowen, D\BPBI A.%
\BDBL {}others%
\end{APACrefauthors}%
\unskip\
\newblock
\APACrefYearMonthDay{2020}{}{}.
\newblock
{\BBOQ}\APACrefatitle {Development of a machine learning model using multiple,
  heterogeneous data sources to estimate weekly {US} suicide fatalities}
  {Development of a machine learning model using multiple, heterogeneous data
  sources to estimate weekly {US} suicide fatalities}.{\BBCQ}
\newblock
\APACjournalVolNumPages{JAMA Network Open}{3}{12}{e2030932--e2030932}.
\PrintBackRefs{\CurrentBib}

\bibitem [\protect \citeauthoryear {%
Clarke%
, Yan%
, Keusch%
\BCBL {}\ \BBA {} Gallagher%
}{%
Clarke%
\ \protect \BOthers {.}}{%
{\protect \APACyear {2015}}%
}]{%
clarke2015impact}
\APACinsertmetastar {%
clarke2015impact}%
\begin{APACrefauthors}%
Clarke, P\BPBI J.%
, Yan, T.%
, Keusch, F.%
\BCBL {}\ \BBA {} Gallagher, N\BPBI A.%
\end{APACrefauthors}%
\unskip\
\newblock
\APACrefYearMonthDay{2015}{}{}.
\newblock
{\BBOQ}\APACrefatitle {The impact of weather on mobility and participation in
  older {US} adults} {The impact of weather on mobility and participation in
  older {US} adults}.{\BBCQ}
\newblock
\APACjournalVolNumPages{American Journal of Public Health}{105}{7}{1489--1494}.
\PrintBackRefs{\CurrentBib}

\bibitem [\protect \citeauthoryear {%
Clayton%
}{%
Clayton%
}{%
{\protect \APACyear {2020}}%
}]{%
clayton2020climate}
\APACinsertmetastar {%
clayton2020climate}%
\begin{APACrefauthors}%
Clayton, S.%
\end{APACrefauthors}%
\unskip\
\newblock
\APACrefYearMonthDay{2020}{}{}.
\newblock
{\BBOQ}\APACrefatitle {Climate anxiety: Psychological responses to climate
  change} {Climate anxiety: Psychological responses to climate change}.{\BBCQ}
\newblock
\APACjournalVolNumPages{Journal of Anxiety Disorders}{74}{}{102263}.
\PrintBackRefs{\CurrentBib}

\bibitem [\protect \citeauthoryear {%
Cusack%
, de Crespigny%
\BCBL {}\ \BBA {} Athanasos%
}{%
Cusack%
\ \protect \BOthers {.}}{%
{\protect \APACyear {2011}}%
}]{%
cusack2011heatwaves}
\APACinsertmetastar {%
cusack2011heatwaves}%
\begin{APACrefauthors}%
Cusack, L.%
, de Crespigny, C.%
\BCBL {}\ \BBA {} Athanasos, P.%
\end{APACrefauthors}%
\unskip\
\newblock
\APACrefYearMonthDay{2011}{}{}.
\newblock
{\BBOQ}\APACrefatitle {Heatwaves and their impact on people with alcohol, drug
  and mental health conditions: A discussion paper on clinical practice
  considerations} {Heatwaves and their impact on people with alcohol, drug and
  mental health conditions: A discussion paper on clinical practice
  considerations}.{\BBCQ}
\newblock
\APACjournalVolNumPages{Journal of Advanced Nursing}{67}{4}{915--922}.
\PrintBackRefs{\CurrentBib}

\bibitem [\protect \citeauthoryear {%
Dell%
, Jones%
\BCBL {}\ \BBA {} Olken%
}{%
Dell%
\ \protect \BOthers {.}}{%
{\protect \APACyear {2012}}%
}]{%
dell2012temperature}
\APACinsertmetastar {%
dell2012temperature}%
\begin{APACrefauthors}%
Dell, M.%
, Jones, B\BPBI F.%
\BCBL {}\ \BBA {} Olken, B\BPBI A.%
\end{APACrefauthors}%
\unskip\
\newblock
\APACrefYearMonthDay{2012}{}{}.
\newblock
{\BBOQ}\APACrefatitle {Temperature shocks and economic growth: Evidence from
  the last half century} {Temperature shocks and economic growth: Evidence from
  the last half century}.{\BBCQ}
\newblock
\APACjournalVolNumPages{American Economic Journal:
  Macroeconomics}{4}{3}{66--95}.
\PrintBackRefs{\CurrentBib}

\bibitem [\protect \citeauthoryear {%
Dell%
, Jones%
\BCBL {}\ \BBA {} Olken%
}{%
Dell%
\ \protect \BOthers {.}}{%
{\protect \APACyear {2014}}%
}]{%
dell2014we}
\APACinsertmetastar {%
dell2014we}%
\begin{APACrefauthors}%
Dell, M.%
, Jones, B\BPBI F.%
\BCBL {}\ \BBA {} Olken, B\BPBI A.%
\end{APACrefauthors}%
\unskip\
\newblock
\APACrefYearMonthDay{2014}{}{}.
\newblock
{\BBOQ}\APACrefatitle {What do we learn from the weather? {The} new
  climate-economy literature} {What do we learn from the weather? {The} new
  climate-economy literature}.{\BBCQ}
\newblock
\APACjournalVolNumPages{Journal of Economic Literature}{52}{3}{740--98}.
\PrintBackRefs{\CurrentBib}

\bibitem [\protect \citeauthoryear {%
Demyttenaere%
\ \protect \BOthers {.}}{%
Demyttenaere%
\ \protect \BOthers {.}}{%
{\protect \APACyear {2004}}%
}]{%
who2004prevalence}
\APACinsertmetastar {%
who2004prevalence}%
\begin{APACrefauthors}%
Demyttenaere, K.%
\BCBT {}\ \BOthersPeriod {.}
\end{APACrefauthors}%
\unskip\
\newblock
\APACrefYearMonthDay{2004}{}{}.
\newblock
{\BBOQ}\APACrefatitle {Prevalence, severity, and unmet need for treatment of
  mental disorders in the World Health Organization World Mental Health
  Surveys} {Prevalence, severity, and unmet need for treatment of mental
  disorders in the world health organization world mental health
  surveys}.{\BBCQ}
\newblock
\APACjournalVolNumPages{Journal of the American Medical
  Association}{291}{21}{2581--2590}.
\PrintBackRefs{\CurrentBib}

\bibitem [\protect \citeauthoryear {%
Deschenes%
}{%
Deschenes%
}{%
{\protect \APACyear {2014}}%
}]{%
deschenes2014temperature}
\APACinsertmetastar {%
deschenes2014temperature}%
\begin{APACrefauthors}%
Deschenes, O.%
\end{APACrefauthors}%
\unskip\
\newblock
\APACrefYearMonthDay{2014}{}{}.
\newblock
{\BBOQ}\APACrefatitle {Temperature, human health, and adaptation: A review of
  the empirical literature} {Temperature, human health, and adaptation: A
  review of the empirical literature}.{\BBCQ}
\newblock
\APACjournalVolNumPages{Energy Economics}{46}{}{606--619}.
\PrintBackRefs{\CurrentBib}

\bibitem [\protect \citeauthoryear {%
Desch{\^e}nes%
\ \BBA {} Greenstone%
}{%
Desch{\^e}nes%
\ \BBA {} Greenstone%
}{%
{\protect \APACyear {2011}}%
}]{%
deschenes2011climate}
\APACinsertmetastar {%
deschenes2011climate}%
\begin{APACrefauthors}%
Desch{\^e}nes, O.%
\BCBT {}\ \BBA {} Greenstone, M.%
\end{APACrefauthors}%
\unskip\
\newblock
\APACrefYearMonthDay{2011}{}{}.
\newblock
{\BBOQ}\APACrefatitle {Climate change, mortality, and adaptation: Evidence from
  annual fluctuations in weather in the US} {Climate change, mortality, and
  adaptation: Evidence from annual fluctuations in weather in the us}.{\BBCQ}
\newblock
\APACjournalVolNumPages{American Economic Journal: Applied
  Economics}{3}{4}{152--85}.
\PrintBackRefs{\CurrentBib}

\bibitem [\protect \citeauthoryear {%
Desch{\^e}nes%
, Greenstone%
\BCBL {}\ \BBA {} Guryan%
}{%
Desch{\^e}nes%
\ \protect \BOthers {.}}{%
{\protect \APACyear {2009}}%
}]{%
deschenes2009climate}
\APACinsertmetastar {%
deschenes2009climate}%
\begin{APACrefauthors}%
Desch{\^e}nes, O.%
, Greenstone, M.%
\BCBL {}\ \BBA {} Guryan, J.%
\end{APACrefauthors}%
\unskip\
\newblock
\APACrefYearMonthDay{2009}{}{}.
\newblock
{\BBOQ}\APACrefatitle {Climate change and birth weight} {Climate change and
  birth weight}.{\BBCQ}
\newblock
\APACjournalVolNumPages{American Economic Review}{99}{2}{211--17}.
\PrintBackRefs{\CurrentBib}

\bibitem [\protect \citeauthoryear {%
Deschenes%
\ \BBA {} Moretti%
}{%
Deschenes%
\ \BBA {} Moretti%
}{%
{\protect \APACyear {2009}}%
}]{%
deschenes2009extreme}
\APACinsertmetastar {%
deschenes2009extreme}%
\begin{APACrefauthors}%
Deschenes, O.%
\BCBT {}\ \BBA {} Moretti, E.%
\end{APACrefauthors}%
\unskip\
\newblock
\APACrefYearMonthDay{2009}{}{}.
\newblock
{\BBOQ}\APACrefatitle {Extreme weather events, mortality, and migration}
  {Extreme weather events, mortality, and migration}.{\BBCQ}
\newblock
\APACjournalVolNumPages{The Review of Economics and
  Statistics}{91}{4}{659--681}.
\PrintBackRefs{\CurrentBib}

\bibitem [\protect \citeauthoryear {%
{DWD}%
}{%
{DWD}%
}{%
{\protect \APACyear {2020}}%
}]{%
DWD2020}
\APACinsertmetastar {%
DWD2020}%
\begin{APACrefauthors}%
{DWD}.%
\end{APACrefauthors}%
\unskip\
\newblock
\APACrefYear{2020}.
\newblock
\APACrefbtitle {{Klimastatusbericht Deutschland Jahr 2019}}
  {{Klimastatusbericht Deutschland Jahr 2019}}.
\newblock
\APACaddressPublisher{}{Deutscher Wetterdienst}.
\PrintBackRefs{\CurrentBib}

\bibitem [\protect \citeauthoryear {%
Fernandez-Mendoza%
\ \BBA {} Vgontzas%
}{%
Fernandez-Mendoza%
\ \BBA {} Vgontzas%
}{%
{\protect \APACyear {2013}}%
}]{%
fernandez2013insomnia}
\APACinsertmetastar {%
fernandez2013insomnia}%
\begin{APACrefauthors}%
Fernandez-Mendoza, J.%
\BCBT {}\ \BBA {} Vgontzas, A\BPBI N.%
\end{APACrefauthors}%
\unskip\
\newblock
\APACrefYearMonthDay{2013}{}{}.
\newblock
{\BBOQ}\APACrefatitle {Insomnia and its impact on physical and mental health}
  {Insomnia and its impact on physical and mental health}.{\BBCQ}
\newblock
\APACjournalVolNumPages{Current Psychiatry Reports}{15}{12}{418}.
\PrintBackRefs{\CurrentBib}

\bibitem [\protect \citeauthoryear {%
Ferrari%
\ \protect \BOthers {.}}{%
Ferrari%
\ \protect \BOthers {.}}{%
{\protect \APACyear {2022}}%
}]{%
alize2022global}
\APACinsertmetastar {%
alize2022global}%
\begin{APACrefauthors}%
Ferrari, A\BPBI J.%
, Santomauro, D\BPBI F.%
, Herrera, A\BPBI M\BPBI M.%
, Ashbaugh, J\BPBI S\BPBI C.%
, Erskine, H\BPBI E.%
, Charlson, F\BPBI J.%
\BDBL {}Whiteford., H\BPBI A.%
\end{APACrefauthors}%
\unskip\
\newblock
\APACrefYearMonthDay{2022}{}{}.
\newblock
{\BBOQ}\APACrefatitle {Global, regional, and national burden of 12 mental
  disorders in 204 countries and territories, 1990–2019: a systematic
  analysis for the Global Burden of Disease Study 2019} {Global, regional, and
  national burden of 12 mental disorders in 204 countries and territories,
  1990–2019: a systematic analysis for the global burden of disease study
  2019}.{\BBCQ}
\newblock
\APACjournalVolNumPages{The Lancet Psychiatry}{}{}{forthcoming}.
\PrintBackRefs{\CurrentBib}

\bibitem [\protect \citeauthoryear {%
Fishman%
, Carrillo%
\BCBL {}\ \BBA {} Russ%
}{%
Fishman%
\ \protect \BOthers {.}}{%
{\protect \APACyear {2019}}%
}]{%
fishman2019long}
\APACinsertmetastar {%
fishman2019long}%
\begin{APACrefauthors}%
Fishman, R.%
, Carrillo, P.%
\BCBL {}\ \BBA {} Russ, J.%
\end{APACrefauthors}%
\unskip\
\newblock
\APACrefYearMonthDay{2019}{}{}.
\newblock
{\BBOQ}\APACrefatitle {Long-term impacts of exposure to high temperatures on
  human capital and economic productivity} {Long-term impacts of exposure to
  high temperatures on human capital and economic productivity}.{\BBCQ}
\newblock
\APACjournalVolNumPages{Journal of Environmental Economics and
  Management}{93}{}{221--238}.
\PrintBackRefs{\CurrentBib}

\bibitem [\protect \citeauthoryear {%
Gardner%
\ \BBA {} Oswald%
}{%
Gardner%
\ \BBA {} Oswald%
}{%
{\protect \APACyear {2007}}%
}]{%
gardner2007money}
\APACinsertmetastar {%
gardner2007money}%
\begin{APACrefauthors}%
Gardner, J.%
\BCBT {}\ \BBA {} Oswald, A\BPBI J.%
\end{APACrefauthors}%
\unskip\
\newblock
\APACrefYearMonthDay{2007}{}{}.
\newblock
{\BBOQ}\APACrefatitle {Money and mental wellbeing: A longitudinal study of
  medium-sized lottery wins} {Money and mental wellbeing: A longitudinal study
  of medium-sized lottery wins}.{\BBCQ}
\newblock
\APACjournalVolNumPages{Journal of Health Economics}{26}{1}{49--60}.
\PrintBackRefs{\CurrentBib}

\bibitem [\protect \citeauthoryear {%
Gathergood%
}{%
Gathergood%
}{%
{\protect \APACyear {2012}}%
}]{%
gathergood2012debt}
\APACinsertmetastar {%
gathergood2012debt}%
\begin{APACrefauthors}%
Gathergood, J.%
\end{APACrefauthors}%
\unskip\
\newblock
\APACrefYearMonthDay{2012}{}{}.
\newblock
{\BBOQ}\APACrefatitle {Debt and depression: causal links and social norm
  effects} {Debt and depression: causal links and social norm effects}.{\BBCQ}
\newblock
\APACjournalVolNumPages{The Economic Journal}{122}{563}{1094--1114}.
\PrintBackRefs{\CurrentBib}

\bibitem [\protect \citeauthoryear {%
Graff~Zivin%
, Hsiang%
\BCBL {}\ \BBA {} Neidell%
}{%
Graff~Zivin%
\ \protect \BOthers {.}}{%
{\protect \APACyear {2018}}%
}]{%
graff2018temperature}
\APACinsertmetastar {%
graff2018temperature}%
\begin{APACrefauthors}%
Graff~Zivin, J.%
, Hsiang, S\BPBI M.%
\BCBL {}\ \BBA {} Neidell, M.%
\end{APACrefauthors}%
\unskip\
\newblock
\APACrefYearMonthDay{2018}{}{}.
\newblock
{\BBOQ}\APACrefatitle {Temperature and human capital in the short and long run}
  {Temperature and human capital in the short and long run}.{\BBCQ}
\newblock
\APACjournalVolNumPages{Journal of the Association of Environmental and
  Resource Economists}{5}{1}{77--105}.
\PrintBackRefs{\CurrentBib}

\bibitem [\protect \citeauthoryear {%
Hansen%
\ \protect \BOthers {.}}{%
Hansen%
\ \protect \BOthers {.}}{%
{\protect \APACyear {2008}}%
}]{%
hansen2008effect}
\APACinsertmetastar {%
hansen2008effect}%
\begin{APACrefauthors}%
Hansen, A.%
, Bi, P.%
, Nitschke, M.%
, Ryan, P.%
, Pisaniello, D.%
\BCBL {}\ \BBA {} Tucker, G.%
\end{APACrefauthors}%
\unskip\
\newblock
\APACrefYearMonthDay{2008}{}{}.
\newblock
{\BBOQ}\APACrefatitle {The effect of heat waves on mental health in a temperate
  {A}ustralian city} {The effect of heat waves on mental health in a temperate
  {A}ustralian city}.{\BBCQ}
\newblock
\APACjournalVolNumPages{Environmental Health
  Perspectives}{116}{10}{1369--1375}.
\PrintBackRefs{\CurrentBib}

\bibitem [\protect \citeauthoryear {%
Heyes%
\ \BBA {} Saberian%
}{%
Heyes%
\ \BBA {} Saberian%
}{%
{\protect \APACyear {2019}}%
}]{%
heyes2019temperature}
\APACinsertmetastar {%
heyes2019temperature}%
\begin{APACrefauthors}%
Heyes, A.%
\BCBT {}\ \BBA {} Saberian, S.%
\end{APACrefauthors}%
\unskip\
\newblock
\APACrefYearMonthDay{2019}{}{}.
\newblock
{\BBOQ}\APACrefatitle {Temperature and decisions: Evidence from 207,000 court
  cases} {Temperature and decisions: Evidence from 207,000 court cases}.{\BBCQ}
\newblock
\APACjournalVolNumPages{American Economic Journal: Applied
  Economics}{11}{2}{238--65}.
\PrintBackRefs{\CurrentBib}

\bibitem [\protect \citeauthoryear {%
Hsiang%
}{%
Hsiang%
}{%
{\protect \APACyear {2016}}%
}]{%
hsiang2016climate}
\APACinsertmetastar {%
hsiang2016climate}%
\begin{APACrefauthors}%
Hsiang, S.%
\end{APACrefauthors}%
\unskip\
\newblock
\APACrefYearMonthDay{2016}{}{}.
\newblock
{\BBOQ}\APACrefatitle {Climate econometrics} {Climate econometrics}.{\BBCQ}
\newblock
\APACjournalVolNumPages{Annual Review of Resource Economics}{8}{}{43--75}.
\PrintBackRefs{\CurrentBib}

\bibitem [\protect \citeauthoryear {%
Hsiang%
, Burke%
\BCBL {}\ \BBA {} Miguel%
}{%
Hsiang%
\ \protect \BOthers {.}}{%
{\protect \APACyear {2013}}%
}]{%
hsiang2013quantifying}
\APACinsertmetastar {%
hsiang2013quantifying}%
\begin{APACrefauthors}%
Hsiang, S.%
, Burke, M.%
\BCBL {}\ \BBA {} Miguel, E.%
\end{APACrefauthors}%
\unskip\
\newblock
\APACrefYearMonthDay{2013}{}{}.
\newblock
{\BBOQ}\APACrefatitle {Quantifying the influence of climate on human conflict}
  {Quantifying the influence of climate on human conflict}.{\BBCQ}
\newblock
\APACjournalVolNumPages{Science}{341}{6151}{1235367}.
\PrintBackRefs{\CurrentBib}

\bibitem [\protect \citeauthoryear {%
IPCC%
}{%
IPCC%
}{%
{\protect \APACyear {2021}}%
}]{%
IPCC2021}
\APACinsertmetastar {%
IPCC2021}%
\begin{APACrefauthors}%
IPCC.%
\end{APACrefauthors}%
\unskip\
\newblock
\APACrefYearMonthDay{2021}{}{}.
\newblock
{\BBOQ}\APACrefatitle {Summary for Policymakers} {Summary for
  policymakers}.{\BBCQ}
\newblock
\BIn{} \APACrefbtitle {Climate Change 2021: The Physical Science Basis.
  Contribution of Working Group {I} to the Sixth Assessment Report of the
  Intergovernmental Panel on Climate Change.} {Climate change 2021: The
  physical science basis. contribution of working group {I} to the sixth
  assessment report of the intergovernmental panel on climate change.}
\newblock
\APACaddressPublisher{}{Cambridge University Press}.
\PrintBackRefs{\CurrentBib}

\bibitem [\protect \citeauthoryear {%
Jahedi%
\ \BBA {} M{\'e}ndez%
}{%
Jahedi%
\ \BBA {} M{\'e}ndez%
}{%
{\protect \APACyear {2014}}%
}]{%
jahedi2014advantages}
\APACinsertmetastar {%
jahedi2014advantages}%
\begin{APACrefauthors}%
Jahedi, S.%
\BCBT {}\ \BBA {} M{\'e}ndez, F.%
\end{APACrefauthors}%
\unskip\
\newblock
\APACrefYearMonthDay{2014}{}{}.
\newblock
{\BBOQ}\APACrefatitle {On the advantages and disadvantages of subjective
  measures} {On the advantages and disadvantages of subjective
  measures}.{\BBCQ}
\newblock
\APACjournalVolNumPages{Journal of Economic Behavior \&
  Organization}{98}{}{97--114}.
\PrintBackRefs{\CurrentBib}

\bibitem [\protect \citeauthoryear {%
Jin%
\ \BBA {} Ziebarth%
}{%
Jin%
\ \BBA {} Ziebarth%
}{%
{\protect \APACyear {2020}}%
}]{%
jin2020sleep}
\APACinsertmetastar {%
jin2020sleep}%
\begin{APACrefauthors}%
Jin, L.%
\BCBT {}\ \BBA {} Ziebarth, N\BPBI R.%
\end{APACrefauthors}%
\unskip\
\newblock
\APACrefYearMonthDay{2020}{}{}.
\newblock
{\BBOQ}\APACrefatitle {Sleep, health, and human capital: Evidence from daylight
  saving time} {Sleep, health, and human capital: Evidence from daylight saving
  time}.{\BBCQ}
\newblock
\APACjournalVolNumPages{Journal of Economic Behavior \&
  Organization}{170}{}{174--192}.
\PrintBackRefs{\CurrentBib}

\bibitem [\protect \citeauthoryear {%
Karlsson%
\ \BBA {} Ziebarth%
}{%
Karlsson%
\ \BBA {} Ziebarth%
}{%
{\protect \APACyear {2018}}%
}]{%
karlsson2018population}
\APACinsertmetastar {%
karlsson2018population}%
\begin{APACrefauthors}%
Karlsson, M.%
\BCBT {}\ \BBA {} Ziebarth, N\BPBI R.%
\end{APACrefauthors}%
\unskip\
\newblock
\APACrefYearMonthDay{2018}{}{}.
\newblock
{\BBOQ}\APACrefatitle {Population health effects and health-related costs of
  extreme temperatures: Comprehensive evidence from Germany} {Population health
  effects and health-related costs of extreme temperatures: Comprehensive
  evidence from germany}.{\BBCQ}
\newblock
\APACjournalVolNumPages{Journal of Environmental Economics and
  Management}{91}{}{93--117}.
\PrintBackRefs{\CurrentBib}

\bibitem [\protect \citeauthoryear {%
Katz%
, Kling%
\BCBL {}\ \BBA {} Liebman%
}{%
Katz%
\ \protect \BOthers {.}}{%
{\protect \APACyear {2001}}%
}]{%
katz2001moving}
\APACinsertmetastar {%
katz2001moving}%
\begin{APACrefauthors}%
Katz, L\BPBI F.%
, Kling, J\BPBI R.%
\BCBL {}\ \BBA {} Liebman, J\BPBI B.%
\end{APACrefauthors}%
\unskip\
\newblock
\APACrefYearMonthDay{2001}{}{}.
\newblock
{\BBOQ}\APACrefatitle {Moving to opportunity in Boston: Early results of a
  randomized mobility experiment} {Moving to opportunity in boston: Early
  results of a randomized mobility experiment}.{\BBCQ}
\newblock
\APACjournalVolNumPages{The Quarterly Journal of Economics}{116}{2}{607--654}.
\PrintBackRefs{\CurrentBib}

\bibitem [\protect \citeauthoryear {%
Kiessling%
\ \BBA {} Norris%
}{%
Kiessling%
\ \BBA {} Norris%
}{%
{\protect \APACyear {2022}}%
}]{%
kiessling2022long}
\APACinsertmetastar {%
kiessling2022long}%
\begin{APACrefauthors}%
Kiessling, L.%
\BCBT {}\ \BBA {} Norris, J.%
\end{APACrefauthors}%
\unskip\
\newblock
\APACrefYearMonthDay{2022}{}{}.
\newblock
{\BBOQ}\APACrefatitle {The long-run effects of peers on mental health} {The
  long-run effects of peers on mental health}.{\BBCQ}
\newblock
\APACjournalVolNumPages{The Economic Journal}{}{}{forthcoming}.
\PrintBackRefs{\CurrentBib}

\bibitem [\protect \citeauthoryear {%
Leslie%
\ \BBA {} Wilson%
}{%
Leslie%
\ \BBA {} Wilson%
}{%
{\protect \APACyear {2020}}%
}]{%
leslie2020sheltering}
\APACinsertmetastar {%
leslie2020sheltering}%
\begin{APACrefauthors}%
Leslie, E.%
\BCBT {}\ \BBA {} Wilson, R.%
\end{APACrefauthors}%
\unskip\
\newblock
\APACrefYearMonthDay{2020}{}{}.
\newblock
{\BBOQ}\APACrefatitle {Sheltering in place and domestic violence: Evidence from
  calls for service during {COVID-19}} {Sheltering in place and domestic
  violence: Evidence from calls for service during {COVID-19}}.{\BBCQ}
\newblock
\APACjournalVolNumPages{Journal of Public Economics}{189}{}{104241}.
\PrintBackRefs{\CurrentBib}

\bibitem [\protect \citeauthoryear {%
Liu%
\ \BBA {} Tsai%
}{%
Liu%
\ \BBA {} Tsai%
}{%
{\protect \APACyear {2021}}%
}]{%
liu2021helpline}
\APACinsertmetastar {%
liu2021helpline}%
\begin{APACrefauthors}%
Liu%
\BCBT {}\ \BBA {} Tsai, A.%
\end{APACrefauthors}%
\unskip\
\newblock
\APACrefYearMonthDay{2021}{}{}.
\newblock
{\BBOQ}\APACrefatitle {Helpline data used to monitor population distress in a
  pandemic} {Helpline data used to monitor population distress in a
  pandemic}.{\BBCQ}
\newblock
\APACjournalVolNumPages{Nature}{600}{}{46-47}.
\PrintBackRefs{\CurrentBib}

\bibitem [\protect \citeauthoryear {%
L{\~o}hmus%
}{%
L{\~o}hmus%
}{%
{\protect \APACyear {2018}}%
}]{%
lohmus2018possible}
\APACinsertmetastar {%
lohmus2018possible}%
\begin{APACrefauthors}%
L{\~o}hmus, M.%
\end{APACrefauthors}%
\unskip\
\newblock
\APACrefYearMonthDay{2018}{}{}.
\newblock
{\BBOQ}\APACrefatitle {Possible biological mechanisms linking mental health and
  heat - {A} contemplative review} {Possible biological mechanisms linking
  mental health and heat - {A} contemplative review}.{\BBCQ}
\newblock
\APACjournalVolNumPages{International Journal of Environmental Research and
  Public Health}{15}{7}{1515}.
\PrintBackRefs{\CurrentBib}

\bibitem [\protect \citeauthoryear {%
LoPalo%
}{%
LoPalo%
}{%
{\protect \APACyear {2022}}%
}]{%
lopalo2022temperature}
\APACinsertmetastar {%
lopalo2022temperature}%
\begin{APACrefauthors}%
LoPalo, M.%
\end{APACrefauthors}%
\unskip\
\newblock
\APACrefYearMonthDay{2022}{}{}.
\newblock
{\BBOQ}\APACrefatitle {Temperature, worker productivity, and adaptation:
  Evidence from survey data production} {Temperature, worker productivity, and
  adaptation: Evidence from survey data production}.{\BBCQ}
\newblock
\APACjournalVolNumPages{American Economic Journal: Applied
  Economics}{}{}{forthcoming}.
\PrintBackRefs{\CurrentBib}

\bibitem [\protect \citeauthoryear {%
Massazza%
, Teyton%
, Charlson%
, Benmarhnia%
\BCBL {}\ \BBA {} Augustinavicius%
}{%
Massazza%
\ \protect \BOthers {.}}{%
{\protect \APACyear {2022}}%
}]{%
massazza2022quantitative}
\APACinsertmetastar {%
massazza2022quantitative}%
\begin{APACrefauthors}%
Massazza, A.%
, Teyton, A.%
, Charlson, F.%
, Benmarhnia, T.%
\BCBL {}\ \BBA {} Augustinavicius, J\BPBI L.%
\end{APACrefauthors}%
\unskip\
\newblock
\APACrefYearMonthDay{2022}{}{}.
\newblock
{\BBOQ}\APACrefatitle {Quantitative methods for climate change and mental
  health research: current trends and future directions} {Quantitative methods
  for climate change and mental health research: current trends and future
  directions}.{\BBCQ}
\newblock
\APACjournalVolNumPages{The Lancet Planetary Health}{6}{7}{e613--e627}.
\PrintBackRefs{\CurrentBib}

\bibitem [\protect \citeauthoryear {%
McMichael%
, Woodruff%
\BCBL {}\ \BBA {} Hales%
}{%
McMichael%
\ \protect \BOthers {.}}{%
{\protect \APACyear {2006}}%
}]{%
mcmichael2006climate}
\APACinsertmetastar {%
mcmichael2006climate}%
\begin{APACrefauthors}%
McMichael, A\BPBI J.%
, Woodruff, R\BPBI E.%
\BCBL {}\ \BBA {} Hales, S.%
\end{APACrefauthors}%
\unskip\
\newblock
\APACrefYearMonthDay{2006}{}{}.
\newblock
{\BBOQ}\APACrefatitle {Climate change and human health: Present and future
  risks} {Climate change and human health: Present and future risks}.{\BBCQ}
\newblock
\APACjournalVolNumPages{The Lancet}{367}{9513}{859--869}.
\PrintBackRefs{\CurrentBib}

\bibitem [\protect \citeauthoryear {%
Minor%
, Bjerre-Nielsen%
, Jonasdottir%
, Lehmann%
\BCBL {}\ \BBA {} Obradovich%
}{%
Minor%
\ \protect \BOthers {.}}{%
{\protect \APACyear {2022}}%
}]{%
minor2022rising}
\APACinsertmetastar {%
minor2022rising}%
\begin{APACrefauthors}%
Minor, K.%
, Bjerre-Nielsen, A.%
, Jonasdottir, S\BPBI S.%
, Lehmann, S.%
\BCBL {}\ \BBA {} Obradovich, N.%
\end{APACrefauthors}%
\unskip\
\newblock
\APACrefYearMonthDay{2022}{}{}.
\newblock
{\BBOQ}\APACrefatitle {Rising temperatures erode human sleep globally} {Rising
  temperatures erode human sleep globally}.{\BBCQ}
\newblock
\APACjournalVolNumPages{One Earth}{5}{5}{534--549}.
\PrintBackRefs{\CurrentBib}

\bibitem [\protect \citeauthoryear {%
Mukherjee%
\ \BBA {} Sanders%
}{%
Mukherjee%
\ \BBA {} Sanders%
}{%
{\protect \APACyear {2021}}%
}]{%
mukherjee2021causal}
\APACinsertmetastar {%
mukherjee2021causal}%
\begin{APACrefauthors}%
Mukherjee, A.%
\BCBT {}\ \BBA {} Sanders, N\BPBI J.%
\end{APACrefauthors}%
\unskip\
\newblock
\APACrefYearMonthDay{2021}{}{}.
\newblock
{\BBOQ}\APACrefatitle {The causal effect of heat on violence: Social
  implications of unmitigated heat among the incarcerated} {The causal effect
  of heat on violence: Social implications of unmitigated heat among the
  incarcerated}.{\BBCQ}
\newblock
\APACjournalVolNumPages{NBER Working Paper No. 28987}{}{}{}.
\PrintBackRefs{\CurrentBib}

\bibitem [\protect \citeauthoryear {%
Mullins%
\ \BBA {} White%
}{%
Mullins%
\ \BBA {} White%
}{%
{\protect \APACyear {2019}}%
}]{%
mullins2019temperature}
\APACinsertmetastar {%
mullins2019temperature}%
\begin{APACrefauthors}%
Mullins, J\BPBI T.%
\BCBT {}\ \BBA {} White, C.%
\end{APACrefauthors}%
\unskip\
\newblock
\APACrefYearMonthDay{2019}{}{}.
\newblock
{\BBOQ}\APACrefatitle {Temperature and mental health: Evidence from the
  spectrum of mental health outcomes} {Temperature and mental health: Evidence
  from the spectrum of mental health outcomes}.{\BBCQ}
\newblock
\APACjournalVolNumPages{Journal of Health Economics}{68}{}{102240}.
\PrintBackRefs{\CurrentBib}

\bibitem [\protect \citeauthoryear {%
Nori-Sarma%
\ \protect \BOthers {.}}{%
Nori-Sarma%
\ \protect \BOthers {.}}{%
{\protect \APACyear {2022}}%
}]{%
nori2022association}
\APACinsertmetastar {%
nori2022association}%
\begin{APACrefauthors}%
Nori-Sarma, A.%
, Sun, S.%
, Sun, Y.%
, Spangler, K\BPBI R.%
, Oblath, R.%
, Galea, S.%
\BDBL {}Wellenius, G\BPBI A.%
\end{APACrefauthors}%
\unskip\
\newblock
\APACrefYearMonthDay{2022}{}{}.
\newblock
{\BBOQ}\APACrefatitle {Association between ambient heat and risk of emergency
  department visits for mental health among {US} adults, 2010 to 2019}
  {Association between ambient heat and risk of emergency department visits for
  mental health among {US} adults, 2010 to 2019}.{\BBCQ}
\newblock
\APACjournalVolNumPages{JAMA psychiatry}{}{}{}.
\PrintBackRefs{\CurrentBib}

\bibitem [\protect \citeauthoryear {%
Obradovich%
, Migliorini%
, Mednick%
\BCBL {}\ \BBA {} Fowler%
}{%
Obradovich%
\ \protect \BOthers {.}}{%
{\protect \APACyear {2017}}%
}]{%
obradovich2017nighttime}
\APACinsertmetastar {%
obradovich2017nighttime}%
\begin{APACrefauthors}%
Obradovich, N.%
, Migliorini, R.%
, Mednick, S\BPBI C.%
\BCBL {}\ \BBA {} Fowler, J\BPBI H.%
\end{APACrefauthors}%
\unskip\
\newblock
\APACrefYearMonthDay{2017}{}{}.
\newblock
{\BBOQ}\APACrefatitle {Nighttime temperature and human sleep loss in a changing
  climate} {Nighttime temperature and human sleep loss in a changing
  climate}.{\BBCQ}
\newblock
\APACjournalVolNumPages{Science Advances}{3}{5}{e1601555}.
\PrintBackRefs{\CurrentBib}

\bibitem [\protect \citeauthoryear {%
Obradovich%
, Migliorini%
, Paulus%
\BCBL {}\ \BBA {} Rahwan%
}{%
Obradovich%
\ \protect \BOthers {.}}{%
{\protect \APACyear {2018}}%
}]{%
obradovich2018empirical}
\APACinsertmetastar {%
obradovich2018empirical}%
\begin{APACrefauthors}%
Obradovich, N.%
, Migliorini, R.%
, Paulus, M\BPBI P.%
\BCBL {}\ \BBA {} Rahwan, I.%
\end{APACrefauthors}%
\unskip\
\newblock
\APACrefYearMonthDay{2018}{}{}.
\newblock
{\BBOQ}\APACrefatitle {Empirical evidence of mental health risks posed by
  climate change} {Empirical evidence of mental health risks posed by climate
  change}.{\BBCQ}
\newblock
\APACjournalVolNumPages{Proceedings of the National Academy of
  Sciences}{115}{43}{10953--10958}.
\PrintBackRefs{\CurrentBib}

\bibitem [\protect \citeauthoryear {%
Obradovich%
\ \BBA {} Minor%
}{%
Obradovich%
\ \BBA {} Minor%
}{%
{\protect \APACyear {2022}}%
}]{%
obradovich2022identifying}
\APACinsertmetastar {%
obradovich2022identifying}%
\begin{APACrefauthors}%
Obradovich, N.%
\BCBT {}\ \BBA {} Minor, K.%
\end{APACrefauthors}%
\unskip\
\newblock
\APACrefYearMonthDay{2022}{}{}.
\newblock
{\BBOQ}\APACrefatitle {Identifying and Preparing for the Mental Health Burden
  of Climate Change} {Identifying and preparing for the mental health burden of
  climate change}.{\BBCQ}
\newblock
\APACjournalVolNumPages{JAMA Psychiatry}{}{}{}.
\PrintBackRefs{\CurrentBib}

\bibitem [\protect \citeauthoryear {%
Park%
}{%
Park%
}{%
{\protect \APACyear {2020}}%
}]{%
park2020hot}
\APACinsertmetastar {%
park2020hot}%
\begin{APACrefauthors}%
Park, R\BPBI J.%
\end{APACrefauthors}%
\unskip\
\newblock
\APACrefYearMonthDay{2020}{}{}.
\newblock
{\BBOQ}\APACrefatitle {Hot temperature and high stakes performance} {Hot
  temperature and high stakes performance}.{\BBCQ}
\newblock
\APACjournalVolNumPages{Journal of Human Resources}{57}{2}{400--434}.
\PrintBackRefs{\CurrentBib}

\bibitem [\protect \citeauthoryear {%
Persson%
\ \BBA {} Rossin-Slater%
}{%
Persson%
\ \BBA {} Rossin-Slater%
}{%
{\protect \APACyear {2018}}%
}]{%
persson2018family}
\APACinsertmetastar {%
persson2018family}%
\begin{APACrefauthors}%
Persson, P.%
\BCBT {}\ \BBA {} Rossin-Slater, M.%
\end{APACrefauthors}%
\unskip\
\newblock
\APACrefYearMonthDay{2018}{}{}.
\newblock
{\BBOQ}\APACrefatitle {Family ruptures, stress, and the mental health of the
  next generation} {Family ruptures, stress, and the mental health of the next
  generation}.{\BBCQ}
\newblock
\APACjournalVolNumPages{American Economic Review}{108}{4-5}{1214--52}.
\PrintBackRefs{\CurrentBib}

\bibitem [\protect \citeauthoryear {%
Prince%
\ \protect \BOthers {.}}{%
Prince%
\ \protect \BOthers {.}}{%
{\protect \APACyear {2007}}%
}]{%
prince2007no}
\APACinsertmetastar {%
prince2007no}%
\begin{APACrefauthors}%
Prince, M.%
, Patel, V.%
, Saxena, S.%
, Maj, M.%
, Maselko, J.%
, Phillips, M\BPBI R.%
\BCBL {}\ \BBA {} Rahman, A.%
\end{APACrefauthors}%
\unskip\
\newblock
\APACrefYearMonthDay{2007}{}{}.
\newblock
{\BBOQ}\APACrefatitle {No health without mental health} {No health without
  mental health}.{\BBCQ}
\newblock
\APACjournalVolNumPages{The Lancet}{370}{9590}{859--877}.
\PrintBackRefs{\CurrentBib}

\bibitem [\protect \citeauthoryear {%
Romanello%
\ \protect \BOthers {.}}{%
Romanello%
\ \protect \BOthers {.}}{%
{\protect \APACyear {2021}}%
}]{%
romanello20212021}
\APACinsertmetastar {%
romanello20212021}%
\begin{APACrefauthors}%
Romanello, M.%
, McGushin, A.%
, Di~Napoli, C.%
, Drummond, P.%
, Hughes, N.%
, Jamart, L.%
\BDBL {}others%
\end{APACrefauthors}%
\unskip\
\newblock
\APACrefYearMonthDay{2021}{}{}.
\newblock
{\BBOQ}\APACrefatitle {The 2021 report of the Lancet Countdown on health and
  climate change: Code red for a healthy future} {The 2021 report of the lancet
  countdown on health and climate change: Code red for a healthy
  future}.{\BBCQ}
\newblock
\APACjournalVolNumPages{The Lancet}{398}{10311}{1619--1662}.
\PrintBackRefs{\CurrentBib}

\bibitem [\protect \citeauthoryear {%
Sexton%
, Wang%
\BCBL {}\ \BBA {} Mullins%
}{%
Sexton%
\ \protect \BOthers {.}}{%
{\protect \APACyear {2021}}%
}]{%
sexton2021heat}
\APACinsertmetastar {%
sexton2021heat}%
\begin{APACrefauthors}%
Sexton, S.%
, Wang, Z.%
\BCBL {}\ \BBA {} Mullins, J\BPBI T.%
\end{APACrefauthors}%
\unskip\
\newblock
\APACrefYearMonthDay{2021}{}{}.
\newblock
{\BBOQ}\APACrefatitle {Heat Adaptation and Human Performance in a Warming
  Climate} {Heat adaptation and human performance in a warming climate}.{\BBCQ}
\newblock
\APACjournalVolNumPages{Journal of the Association of Environmental and
  Resource Economists}{9}{1}{}.
\PrintBackRefs{\CurrentBib}

\bibitem [\protect \citeauthoryear {%
Somanathan%
, Somanathan%
, Sudarshan%
\BCBL {}\ \BBA {} Tewari%
}{%
Somanathan%
\ \protect \BOthers {.}}{%
{\protect \APACyear {2021}}%
}]{%
somanathan2021impact}
\APACinsertmetastar {%
somanathan2021impact}%
\begin{APACrefauthors}%
Somanathan, E.%
, Somanathan, R.%
, Sudarshan, A.%
\BCBL {}\ \BBA {} Tewari, M.%
\end{APACrefauthors}%
\unskip\
\newblock
\APACrefYearMonthDay{2021}{}{}.
\newblock
{\BBOQ}\APACrefatitle {The impact of temperature on productivity and labor
  supply: Evidence from Indian manufacturing} {The impact of temperature on
  productivity and labor supply: Evidence from indian manufacturing}.{\BBCQ}
\newblock
\APACjournalVolNumPages{Journal of Political Economy}{129}{6}{1797--1827}.
\PrintBackRefs{\CurrentBib}

\bibitem [\protect \citeauthoryear {%
Telefonseelsorge%
}{%
Telefonseelsorge%
}{%
{\protect \APACyear {2017}}%
}]{%
Telefonseelsorge2017}
\APACinsertmetastar {%
Telefonseelsorge2017}%
\begin{APACrefauthors}%
Telefonseelsorge.%
\end{APACrefauthors}%
\unskip\
\newblock
\APACrefYear{2017}.
\newblock
\APACrefbtitle {Gesamtstatistik für das {J}ahr 2016} {Gesamtstatistik für das
  {J}ahr 2016}.
\newblock
\APACaddressPublisher{}{Berlin}.
\PrintBackRefs{\CurrentBib}

\bibitem [\protect \citeauthoryear {%
Thompson%
, Berenbaum%
\BCBL {}\ \BBA {} Bredemeier%
}{%
Thompson%
\ \protect \BOthers {.}}{%
{\protect \APACyear {2011}}%
}]{%
thompson2011cross}
\APACinsertmetastar {%
thompson2011cross}%
\begin{APACrefauthors}%
Thompson, R\BPBI J.%
, Berenbaum, H.%
\BCBL {}\ \BBA {} Bredemeier, K.%
\end{APACrefauthors}%
\unskip\
\newblock
\APACrefYearMonthDay{2011}{}{}.
\newblock
{\BBOQ}\APACrefatitle {Cross-sectional and longitudinal relations between
  affective instability and depression} {Cross-sectional and longitudinal
  relations between affective instability and depression}.{\BBCQ}
\newblock
\APACjournalVolNumPages{Journal of Affective Disorders}{130}{1-2}{53--59}.
\PrintBackRefs{\CurrentBib}

\bibitem [\protect \citeauthoryear {%
White%
}{%
White%
}{%
{\protect \APACyear {2017}}%
}]{%
white2017dynamic}
\APACinsertmetastar {%
white2017dynamic}%
\begin{APACrefauthors}%
White, C.%
\end{APACrefauthors}%
\unskip\
\newblock
\APACrefYearMonthDay{2017}{}{}.
\newblock
{\BBOQ}\APACrefatitle {The dynamic relationship between temperature and
  morbidity} {The dynamic relationship between temperature and
  morbidity}.{\BBCQ}
\newblock
\APACjournalVolNumPages{Journal of the Association of Environmental and
  Resource Economists}{4}{4}{1155--1198}.
\PrintBackRefs{\CurrentBib}

\bibitem [\protect \citeauthoryear {%
{World Health Organization}%
}{%
{World Health Organization}%
}{%
{\protect \APACyear {2012}}%
}]{%
WHA2012}
\APACinsertmetastar {%
WHA2012}%
\begin{APACrefauthors}%
{World Health Organization}.%
\end{APACrefauthors}%
\unskip\
\newblock
\APACrefYear{2012}.
\newblock
\APACrefbtitle {Global burden of mental disorders and the need for a
  comprehensive, coordinated response from health and social sectors at the
  country level: report by the Secretariat} {Global burden of mental disorders
  and the need for a comprehensive, coordinated response from health and social
  sectors at the country level: report by the secretariat}.
\newblock
\APACaddressPublisher{}{World Health Assembly, 65.}
\PrintBackRefs{\CurrentBib}

\bibitem [\protect \citeauthoryear {%
Zhang%
, Zhang%
\BCBL {}\ \BBA {} Chen%
}{%
Zhang%
\ \protect \BOthers {.}}{%
{\protect \APACyear {2017}}%
}]{%
zhang2017happiness}
\APACinsertmetastar {%
zhang2017happiness}%
\begin{APACrefauthors}%
Zhang, X.%
, Zhang, X.%
\BCBL {}\ \BBA {} Chen, X.%
\end{APACrefauthors}%
\unskip\
\newblock
\APACrefYearMonthDay{2017}{}{}.
\newblock
{\BBOQ}\APACrefatitle {Happiness in the air: How does a dirty sky affect mental
  health and subjective well-being?} {Happiness in the air: How does a dirty
  sky affect mental health and subjective well-being?}{\BBCQ}
\newblock
\APACjournalVolNumPages{Journal of Environmental Economics and
  Management}{85}{}{81--94}.
\PrintBackRefs{\CurrentBib}

\end{thebibliography}

\clearpage

\appendix

\counterwithin{figure}{section}
\counterwithin{table}{section}

\section{Appendix}

\begin{table}[h!]
\centering
\caption{Summary of sampled helpline crisis centers}
\label{tab:summary_centers}
\scriptsize
\begin{tabularx}{\textwidth}{llYYYY}
\toprule
Location               & Observation period      & No. of days & No. of calls & Avg. calls/day & Std. calls/day \\ \midrule
Aachen                 & 31/12/2018 - 29/02/2020 & 421         & 7,632        & 18.13          & 5.58           \\
Aschaffenburg          & 31/12/2018 - 29/02/2020 & 426         & 10,924       & 25.69          & 4.46           \\
Auerbach               & 17/12/2018 - 29/02/2020 & 410         & 6,776        & 16.53          & 7.10           \\
Bad Oeynhausen         & 01/12/2018 - 29/02/2020 & 456         & 10,599       & 23.24          & 4.40           \\
Bamberg                & 16/09/2019 - 29/02/2020 & 167         & 4,286        & 25.66          & 5.04           \\
Bayreuth               & 01/01/2019 - 29/02/2020 & 418         & 7,153        & 17.11          & 7.17           \\
Berlin                 & 31/12/2018 - 29/02/2020 & 423         & 7,610        & 17.99          & 7.57           \\
Bielefeld              & 15/01/2019 - 29/02/2020 & 401         & 10,442       & 26.04          & 6.45           \\
Bochum                 & 03/11/2018 - 29/02/2020 & 483         & 9,883        & 20.46          & 4.94           \\
Bremen                 & 01/02/2019 - 29/02/2020 & 391         & 9,505        & 24.31          & 7.08           \\
Darmstadt              & 02/01/2019 - 29/02/2020 & 424         & 12,139       & 28.63          & 4.55           \\
Dessau                 & 01/01/2019 - 29/02/2020 & 425         & 8,943        & 21.04          & 5.34           \\
Dortmund               & 21/02/2019 - 29/02/2020 & 366         & 7,267        & 19.86          & 5.99           \\
Erlangen               & 14/01/2019 - 29/02/2020 & 412         & 9,795        & 23.77          & 7.86           \\
Essen (ev.)              & 31/03/2019 - 29/02/2020 & 276         & 5,928        & 21.48          & 5.49           \\
Essen (kath.)                 & 01/01/2019 - 29/02/2020 & 425         & 13,443       & 31.63          & 10.33          \\
Frankfurt a. M. (ev.)   & 01/01/2019 - 29/02/2020 & 425         & 11,415       & 26.86          & 5.19           \\
Frankfurt a. M. (kath.) & 23/06/2019 - 29/02/2020 & 336         & 6,913        & 20.57          & 6.85           \\
Freiburg               & 21/12/2018 - 29/02/2020 & 428         & 14,263       & 33.32          & 7.21           \\
Gießen                 & 04/01/2019 - 29/02/2020 & 381         & 7,981        & 20.95          & 5.97           \\
Greifswald             & 18/12/2018 - 29/02/2020 & 405         & 4,607        & 11.38          & 6.70           \\
Hagen                  & 10/11/2018 - 29/02/2020 & 437         & 8,119        & 18.58          & 4.98           \\
Hamburg (ev.)           & 01/01/2019 - 29/02/2020 & 425         & 13,986       & 32.91          & 8.21           \\
Hamburg (kath.)         & 10/01/2019 - 29/02/2020 & 363         & 4,314        & 11.88          & 6.06           \\
Hamm                   & 02/12/2018 - 29/02/2020 & 444         & 10,479       & 23.60          & 5.06           \\
Hanau                  & 27/11/2018 - 29/02/2020 & 460         & 10,482       & 22.79          & 4.11           \\
Hannover               & 01/07/2019 - 29/02/2020 & 244         & 7,322        & 30.01          & 4.91           \\
Ingolstadt             & 27/02/2019 - 29/02/2020 & 368         & 9,894        & 26.89          & 5.36           \\
Kaiserslautern         & 05/11/2018 - 29/02/2020 & 471         & 11,699       & 24.84          & 6.09           \\
Kassel                 & 16/01/2019 - 29/02/2020 & 395         & 7,489        & 18.96          & 5.69           \\
Krefeld                & 20/12/2018 - 29/02/2020 & 430         & 9,243        & 21.50          & 7.70           \\
Konstanz               & 01/01/2019 - 29/02/2020 & 424         & 8,936        & 21.08          & 6.68           \\
Mainz-Wiesbaden        & 16/10/2019 - 29/02/2020 & 122         & 3,243        & 26.58          & 7.89           \\
Münster                & 16/11/2018 - 29/02/2020 & 454         & 12,818       & 28.23          & 7.13           \\
Neubrandenburg         & 22/12/2018 - 29/02/2020 & 408         & 5,213        & 12.78          & 6.25           \\
Neuss                  & 01/07/2019 - 29/02/2020 & 244         & 5,002        & 20.50          & 5.19           \\
Offenburg              & 20/12/2018 - 29/02/2020 & 425         & 9,285        & 21.85          & 6.92           \\
Pforzheim              & 01/01/2019 - 29/02/2020 & 418         & 12,076       & 28.89          & 5.41           \\
Paderborn              & 02/12/2018 - 29/02/2020 & 450         & 8,333        & 18.52          & 3.90           \\
Passau                 & 31/05/2019 - 29/02/2020 & 274         & 6,327        & 23.09          & 5.28           \\
Recklinghausen         & 10/12/2018 - 29/02/2020 & 445         & 11,162       & 25.08          & 4.77           \\
Rosenheim              & 13/04/2019 - 29/02/2020 & 276         & 3,754        & 13.60          & 7.15           \\
Rostock                & 17/12/2018 - 29/02/2020 & 428         & 9,527        & 22.26          & 6.14           \\
Schwerin               & 01/01/2019 - 29/02/2020 & 425         & 14,142       & 33.28          & 6.56           \\
Siegen                 & 04/02/2019 - 29/02/2020 & 386         & 7,289        & 18.88          & 5.06           \\
Solingen               & 26/12/2018 - 29/02/2020 & 420         & 4,560        & 10.86          & 4.87           \\
Stuttgart (kath.)      & 01/03/2019 - 29/02/2020 & 352         & 12,652       & 35.94          & 9.67           \\
Sylt                   & 03/09/2019 - 29/02/2020 & 151         & 1,270        & 8.41           & 2.35           \\
Tübingen               & 27/09/2019 - 29/02/2020 & 155         & 4,300        & 27.74          & 5.65           \\
Ulm                    & 01/01/2019 - 29/02/2020 & 425         & 13,234       & 31.14          & 4.68           \\
Wehr                   & 01/01/2019 - 29/02/2020 & 420         & 5,475        & 13.04          & 4.44           \\
Weiden                 & 23/12/2018 - 29/02/2020 & 427         & 12,461       & 29.18          & 5.58           \\
Wolfsburg              & 01/01/2019 - 29/02/2020 & 424         & 10,059       & 23.72          & 5.85           \\
Wuppertal              & 01/01/2019 - 29/02/2020 & 425         & 10,022       & 23.58          & 5.20           \\
Würzburg               & 01/01/2019 - 29/02/2020 & 424         & 13,585       & 32.04          & 6.54           \\ \midrule
55 crisis center               &  &         & 485,274       & 22.96          & 8.67           \\ \bottomrule
\end{tabularx}%
\end{table}

\begin{table}[]
\centering
\caption{Classification of call topics}
\label{tab:calltopic}
\scriptsize
\begin{tabularx}{\textwidth}{lYYYYY}
\toprule
Main Topics         & \multicolumn{5}{l}{Individual Topics}                                                                                                                                                                                                      \\ \midrule
Physical well-being & \multicolumn{5}{l}{Physical well-being (discomfort, illnesses, disabilities)}                                                                                                                                                                                                    \\
Psychological well-being   & \multicolumn{5}{l}{Mood, stress, self-harm, self-image, suicidality, other psychological well-being}                                                                                                                                  \\
Violence         & \multicolumn{5}{l}{Physical violence and mental violence, sexual violence}                                                                                                                                                                                     \\
Social well-being      & \multicolumn{5}{l}{\begin{tabular}[c]{@{}l@{}}Loneliness, dating, partnership, parenthood, sexuality, pregnancy, family relations, \\ everyday relations, virtual relations, seperation, integration, carework, death, grief\end{tabular}} \\
Livelihood          & \multicolumn{5}{l}{Work, poverty, finances, housing, school, volunteering}   \\               Other       & \multicolumn{5}{l}{Faith and values, culture, religion, current topics, other topics}           \\ \bottomrule
\end{tabularx}
\end{table}

\begin{landscape}
\begin{table}[]
\centering
\caption{Summary statistics of helpline calls}
\label{tab:summary_calls}
\scriptsize
\begin{tabular}{@{}lcccccccccccc@{}}
\toprule
{\ul Panel A: Caller characteristics} & \multicolumn{2}{c|}{Gender}              & \multicolumn{3}{c|}{Age group}                                & \multicolumn{3}{c|}{Employment status}                                    & \multicolumn{4}{c}{Living situation}            \\
                                      & \multicolumn{1}{c|}{female}    & male    & \multicolumn{1}{c|}{30-} & \multicolumn{1}{c|}{30-50} & 50+   & \multicolumn{1}{c|}{working} & \multicolumn{1}{c|}{non-working} & retired & \multicolumn{2}{c|}{alone} & \multicolumn{2}{c}{not alone}        \\
Percentage share of callers           & 67.35                          & 32.15   & 8.76                     & 28.45                      & 55.54 & 26.17                        & 33.99                            & 20.48   &     \multicolumn{2}{c}{60.88}                   &      \multicolumn{2}{c}{29.50}               \\
                                      &                                &         &                          &                            &       &                              &                                  &         &                            &           &             &             \\
{\ul Panel B: Main topics}            & \multicolumn{2}{c|}{Physical well-being} & \multicolumn{2}{c|}{Psychological well-being}                & \multicolumn{2}{c|}{Violence}        & \multicolumn{2}{c|}{Social well-being}          & \multicolumn{2}{c|}{Livelihood}        & \multicolumn{2}{c}{Other} \\\\[-1.8ex] 
Percentage share of calls             & \multicolumn{2}{c}{18.15}                & \multicolumn{2}{c}{42.42}                             & \multicolumn{2}{c}{2.60}             & \multicolumn{2}{c}{90.80}                  & \multicolumn{2}{c}{14.96}              & \multicolumn{2}{c}{6.35}  \\\\[-1.8ex]  \bottomrule
\multicolumn{12}{l}{Notes: Percentages do not add up to 100. Panel A: As \textit{TelefonSeelsorge} offers anonymous telephone counseling, the information about the callers} \\ 
\multicolumn{12}{l}{is based on  the  counselor’s  assessment. Sometimes counselors do not record information on the caller.  Panel B: Counselors can choose up to three } \\ 
\multicolumn{12}{l}{topics for each call. Therefore some calls have three different topics.} \\ 
\multicolumn{11}{l}{} \\ 
\end{tabular}
\end{table}
\end{landscape}

\clearpage

\begin{table}[b!] \centering 
  \caption{Effect of average daily temperature on helpline call volume, other specifications} 
 \label{tab:results_robust} 
\scriptsize
\begin{tabularx}{\textwidth}{lYYYY}
\\[-1.8ex]\hline 
\hline \\[-1.8ex] 
 & \multicolumn{4}{c}{\textit{Dependent variable:}} \\ 
\cline{2-5} 
\\[-1.8ex] & \multicolumn{4}{c}{helpline calls} \\ 
\\[-1.8ex] & Logarithm & Raw count & IHS & Poisson \\ 
\\[-1.8ex] & (1) & (2) & (3) & (4)\\ 
\hline \\[-1.8ex] 
 $<$ 0$^{\circ}$C & 0.051$^{***}$ & 0.420$^{***}$ & 0.049$^{***}$ & 0.018$^{***}$ \\ 
  & (0.013)& (0.097) & (0.012) & (0.004) \\    

0$^{\circ}$C - 5$^{\circ}$C& 0.032$^{***}$ & 0.374$^{***}$ & 0.031$^{***}$ & 0.016$^{***}$ \\ 
  & (0.011) & (0.142) & (0.010) & (0.006) \\
  
 5$^{\circ}$C - 10$^{\circ}$C & 0.024$^{**}$ & 0.360$^{**}$ & 0.023$^{**}$ & 0.016$^{**}$ \\ 
  & (0.010) & (0.148) & (0.010) & (0.006) \\ 

 10$^{\circ}$C - 15$^{\circ}$C& Ref. & Ref. & Ref. & Ref.  \\ 
  & &&  &  \\ 

 15$^{\circ}$C - 20$^{\circ}$C & 0.004 & 0.244 & 0.004 & 0.010 \\ 
  & (0.008) & (0.144) & (0.008) & (0.006) \\  

  20$^{\circ}$C - 25$^{\circ}$C& $-$0.001 & 0.143 & $-$0.001 & 0.006 \\ 
  & (0.013) & (0.196) & (0.012) & (0.008) \\ 
  
  $>$ 25$^{\circ}$C  & 0.034$^{***}$ & 0.444$^{**}$ & 0.033$^{***}$ & 0.019$^{***}$ \\ 
  & (0.009) & (0.178) & (0.009) & (0.007) \\    
& & & & \\ 
Observations & 21,138& 21,138 & 21,138 & 21,138 \\ 
Adjusted/Pseudo R$^{2}$ & 0.455 & 0.506 & 0.460 & 0.205 \\ 
\hline \\[-1.8ex] 
Counseling-center fixed effects & Yes& Yes & Yes & Yes\\ 
Year-by-month fixed effects & Yes& Yes & Yes & Yes  \\ 
Environmental controls & Yes& Yes & Yes & Yes  \\ 
Day-of-week fixed effects & Yes& Yes & Yes & Yes \\ 
Holiday fixed effects & Yes& Yes & Yes & Yes \\ 
\hline 
\hline \\[-1.8ex] 
\multicolumn{5}{l}{Notes: The dependent variable is the daily number of answered helpline calls. The sample period is November 3, 2018 to} \\ 
\multicolumn{5}{l}{February 29, 2020. The standard errors in parentheses are two-way clustered at the counseling center and year-month} \\
\multicolumn{5}{l}{level. Column (1) is estimated using a log transformation of the dependent variable. Column (2) is estimated using the} \\
\multicolumn{5}{l}{raw count of answered calls as dependent variable. Column (3) is estimated using the inverse hyperbolic sine (IHS) of the } \\
\multicolumn{5}{l}{dependent variable. Column (4) is estimated using Poisson regression. $^{*}$p$<$0.1;$^{**}$p$<$ 0.05;$^{***}$p$<$0.01.} \\
\end{tabularx} 
\end{table}

\clearpage

\begin{table}[b!] \centering 
  \caption{Effect of average daily temperature on helpline call volume, alternate standard errors} 
 \label{tab:results_robusterrors} 
\scriptsize
\begin{tabularx}{\textwidth}{lYYY}
\\[-1.8ex]\hline 
\hline \\[-1.8ex] 
 & \multicolumn{3}{c}{\textit{Dependent variable:}} \\ 
\cline{2-4} 
\\[-1.8ex] & \multicolumn{3}{c}{helpline calls} \\ 
\\[-1.8ex] & counseling-center x year-month & federal-state x year-month & federal-state x year-month (wild bootstrap) \\ 
\\[-1.8ex] & (1) & (2) & (3)\\ 
\hline \\[-1.8ex] 
 $<$ 0$^{\circ}$C & 0.051$^{***}$ & 0.051$^{**}$ & 0.051$^{**}$ \\ 
  & (0.013) & (0.018) & (0.024) \\    

0$^{\circ}$C - 5$^{\circ}$C& 0.032$^{***}$ &  0.032$^{***}$ &  0.032$^{***}$ \\ 
  & (0.011) & (0.008) & (0.012) \\
  
 5$^{\circ}$C - 10$^{\circ}$C & 0.024$^{**}$ & 0.024$^{***}$ & 0.024$^{***}$ \\ 
  & (0.010) & (0.007) & (0.010) \\

 10$^{\circ}$C - 15$^{\circ}$C& Ref. & Ref. & Ref.  \\ 
  & &  &  \\ 

 15$^{\circ}$C - 20$^{\circ}$C  & 0.004 & 0.004 & 0.004 \\ 
  & (0.008) & (0.013) & (0.013) \\ 

  20$^{\circ}$C - 25$^{\circ}$C& $-$0.001 & $-$0.001& $-$0.001 \\ 
  & (0.013) & (0.008) & (0.012) \\
  
  $>$ 25$^{\circ}$C  & 0.034$^{***}$ & 0.034$^{***}$ & 0.034$^{***}$ \\ 
  & (0.009) & (0.005) & (0.003) \\  
& & &  \\ 
Observations & 21,138 & 21,138 & 21,138 \\ 
Adjusted R$^{2}$ & 0.455 & 0.455 & 0.455 \\ 
\hline \\[-1.8ex] 
Counseling-center fixed effects & Yes & Yes & Yes\\ 
Year-by-month fixed effects & Yes & Yes & Yes  \\ 
Environmental controls & Yes & Yes & Yes  \\ 
Day-of-week fixed effects & Yes & Yes & Yes \\ 
Holiday fixed effects & Yes & Yes & Yes \\ 
\hline 
\hline \\[-1.8ex] 
\multicolumn{4}{l}{Notes: The dependent variable is the natural logarithm of the daily number of answered helpline calls. The sample period} \\ 
\multicolumn{4}{l}{is November 3, 2018 to February 29, 2020. The standard errors in Column (1) are two-way clustered at the counseling} \\
\multicolumn{4}{l}{center and year-month level. The standard errors in Column (2) are two-way clustered at the federal state and year-month} \\
\multicolumn{4}{l}{level. The standard errors in Column (3) are two-way clustered at the federal state and year-month level and are obtained} \\
\multicolumn{4}{l}{by wild bootstrap, using 10,000 replications. $^{*}$p$<$0.1;$^{**}$p$<$0.05;$^{***}$p$<$0.01.} \\
\end{tabularx} 
\end{table}

\begin{table}[t!] 
\captionsetup{justification=centering,margin=1cm}
\centering 
  \caption{Effect of average daily temperature on helpline call volume, environmental factors } 
  \label{tab:results_renv} 
\scriptsize
\begin{tabularx}{\textwidth}{lYYYYY}
\\[-1.8ex]\hline 
\hline \\[-1.8ex] 
 & \multicolumn{5}{c}{\textit{Dependent variable:}} 
 \\
 \cline{2-6} 
 \\[-1.8ex] & \multicolumn{5}{c}{helpline calls} \\
\\[-1.8ex] & (1) & (2) & (3) & (4) & (5)\\ 
\hline \\[-1.8ex] 
 $<$ 0$^{\circ}$C & 0.050$^{***}$ & 0.052$^{***}$ & 0.051$^{***}$ & 0.055$^{***}$ & 0.035$^{***}$ \\ 
  & (0.014) & (0.013) & (0.013) & (0.013) & (0.011)\\  

 0$^{\circ}$C - 5$^{\circ}$C& 0.031$^{***}$ & 0.033$^{***}$ & 0.032$^{***}$ & 0.035$^{***}$ & 0.032$^{***}$\\ 
  & (0.011) & (0.011) & (0.011) & (0.010) & (0.011)\\   

 5$^{\circ}$C - 10$^{\circ}$C  & 0.023$^{**}$ & 0.024$^{**}$ & 0.024$^{**}$ & 0.024$^{**}$ & 0.025$^{**}$ \\ 
  & (0.010) & (0.010) & (0.010) & (0.010) & (0.010)\\  

 10$^{\circ}$C - 15$^{\circ}$C & Ref. & Ref. & Ref. & Ref. & Ref.\\ 
 & & & & & \\ 

  15$^{\circ}$C - 20$^{\circ}$C & 0.004 & 0.004 & 0.004 & 0.004 & 0.004 \\ 
  & (0.008) & (0.008) & (0.009) & (0.009) & (0.009) \\    
  
 20$^{\circ}$C - 25$^{\circ}$C  & $-$0.000 & $-$0.002 & $-$0.001 & $-$0.002 & $-$0.006 \\ 
  & (0.013) & (0.012) & (0.014) & (0.013) & (0.014)\\ 

 $>$ 25$^{\circ}$C  & 0.035$^{***}$ & 0.032$^{***}$ & 0.033$^{***}$ & 0.033$^{***}$ & 0.023$^{***}$ \\ 
  & (0.009) & (0.007) & (0.011) & (0.010) & (0.009) \\ 
  & & & & & \\ 
Observations & 21,138 & 21,138 & 21,138 & 21,138 & 21,138  \\ 
Adjusted R$^{2}$ & 0.455 & 0.455 & 0.455 & 0.455 & 0.455 \\ 
\hline \\[-1.8ex] 
Precipitation & No & Yes & Yes & Yes & Yes\\ 
Relative humidity & Yes & No & Yes & Yes & Yes\\ 
Sunshine duration & Yes & Yes & No & Yes & Yes\\ 
Surface wind speed & Yes & Yes & Yes & No & Yes\\ 
Air pollution & Yes & Yes & Yes & Yes & No\\ 
\hline \\[-1.8ex] 
Counseling-center fixed effects & Yes & Yes & Yes & Yes & Yes \\ 
Year-by-month fixed effects & Yes & Yes & Yes & Yes & Yes \\
Day-of-week fixed effects & Yes & Yes & Yes & Yes & Yes\\ 
Holiday fixed effects & Yes & Yes & Yes & Yes & Yes\\ 
\hline 
\hline \\[-1.8ex] 
\multicolumn{6}{l}{Notes: The dependent variable is the natural logarithm of the daily number of answered helpline calls. The sample period} \\ 
\multicolumn{6}{l}{is November 3, 2018 to February 29, 2020. The standard errors in parentheses are two-way clustered at the counseling} \\ 
\multicolumn{6}{l}{center and year-month level.$^{*}$p$<$0.1; $^{**}$p$<$0.05; $^{***}$p$<$0.01.  } \\
\end{tabularx} 
\end{table}

\begin{table}[t!] \centering 
\captionsetup{justification=centering,margin=1cm}
\centering 
  \caption{Effect of minimum and maximum daily temperature on helpline call volume} 
  \label{tab:results_r2} 
\scriptsize
\begin{tabularx}{\textwidth}{lYYYY}
\\[-1.8ex]\hline 
\hline \\[-1.8ex] 
& \multicolumn{4}{c}{\textit{Dependent variable:}} \\ 
\cline{2-5} 
\\[-1.8ex] & \multicolumn{4}{c}{helpline calls} \\ 
\\[-1.8ex] & (1) & (2) & (3) & (4)\\ 
\hline \\[-1.8ex] 
\underline{\textit{Panel A. Minimum Temperatures}} &  &  &  \\ 
 $<$ -4$^{\circ}$C& 0.065$^{***}$ & 0.068$^{***}$ & 0.068$^{***}$ & 0.070$^{***}$ \\ 
  & (0.023) & (0.025) & (0.022) & (0.024) \\ 
  
 -4$^{\circ}$C - 1$^{\circ}$C& 0.043$^{**}$ & 0.045$^{**}$ & 0.044$^{**}$ & 0.045$^{**}$ \\ 
  & (0.020) & (0.021) & (0.020) & (0.021) \\

 1$^{\circ}$C - 6$^{\circ}$C& 0.032$^{**}$ & 0.032$^{**}$ & 0.032$^{**}$ & 0.032$^{**}$ \\ 
  & (0.015) & (0.016) & (0.015) & (0.015) \\

 6$^{\circ}$C - 11$^{\circ}$C & Ref. & Ref. & Ref. & Ref. \\ 
 & & & & \\  

 11$^{\circ}$C - 16$^{\circ}$C & $-$0.016 & $-$0.017 & $-$0.017 & $-$0.017 \\ 
  & (0.017) & (0.017) & (0.017) & (0.017) \\  

 $>$ 16$^{\circ}$C & 0.060$^{**}$ & 0.063$^{**}$ & 0.061$^{**}$ & 0.064$^{**}$ \\ 
  & (0.028) & (0.028) & (0.028) & (0.028) \\
  & & & &  \\ 

Observations & 21,138 & 21,138 & 21,138 & 21,138 \\ 
Adjusted R$^{2}$   & 0.455 & 0.454 & 0.455 & 0.454 \\ 
  & & & &  \\ 
\underline{\textit{Panel B. Maximum Temperatures}} &  &  &  \\ 
 $<$ 2$^{\circ}$C & 0.043$^{***}$ & 0.043$^{***}$ & 0.045$^{***}$ & 0.045$^{***}$ \\ 
  & (0.012) & (0.013) & (0.013) & (0.014) \\

 2$^{\circ}$C - 12$^{\circ}$C& 0.024$^{***}$ & 0.026$^{***}$ & 0.024$^{***}$ & 0.026$^{***}$ \\ 
  & (0.007) & (0.008) & (0.007) & (0.008) \\  

 12$^{\circ}$C - 22$^{\circ}$C & Ref. & Ref. & Ref. & Ref. \\ 
 & & & & \\  

 22$^{\circ}$C - 32$^{\circ}$C & 0.004 & $-$0.001 & 0.003 & $-$0.002 \\ 
  & (0.008) & (0.008) & (0.008) & (0.008) \\ 

 $>$ 32$^{\circ}$C & 0.025$^{**}$ & 0.022$^{**}$ & 0.025$^{**}$ & 0.023$^{**}$ \\ 
  & (0.012) & (0.010) & (0.012) & (0.010) \\
  & & & &  \\ 
Observations & 21,138 & 21,138 & 21,138 & 21,138 \\ 
Adjusted R$^{2}$   & 0.455 & 0.454 & 0.455 & 0.454 \\ 
\hline \\[-1.8ex] 
Counseling-center fixed effects & Yes & Yes & Yes & Yes\\ 
Year-by-month fixed effects & Yes & Yes & Yes & Yes  \\ 
Environmental controls & Yes & Yes & Yes & Yes  \\
Day-of-week fixed effects & Yes & No & Yes & No\\ 
Holiday fixed effects & Yes & Yes & No & No \\ 
\hline 
\hline \\[-1.8ex] 
\multicolumn{5}{l}{Notes: The dependent variable is the natural logarithm of the daily number of answered helpline calls. The sample period} \\ 
\multicolumn{5}{l}{is November 3, 2018 to February 29, 2020. The standard errors in parentheses are two-way clustered at the counseling} \\
\multicolumn{5}{l}{center and year-month level. $^{*}$p$<$0.1; $^{**}$p$<$0.05; $^{***}$p$<$0.01.} \\
\end{tabularx}
\end{table} 

\clearpage

\clearpage
\begin{table}[t!] 
\captionsetup{justification=centering,margin=1cm}
\centering 
  \caption{Effect of average daily temperature on helpline call volume by gender and age group} 
  \label{tab:results_gendage} 
\scriptsize 
\begin{tabularx}{\textwidth}{@{\extracolsep{4pt}}lYYYYY@{}}
\\[-1.8ex]\hline 
\hline \\[-1.8ex] 
 & \multicolumn{5}{c}{\textit{Dependent variable:}} 
 \\
 \cline{2-6} 
 \\[-1.8ex] & \multicolumn{5}{c}{helpline calls} \\
\\[-1.8ex]  & \multicolumn{2}{c}{gender}&\multicolumn{3}{c}{age group}\\
\cline{2-3}  \cline{4-6} 
\\[-1.8ex] & female & male & 30- & 30-50 & 50+ \\ 
\\[-1.8ex] & (1) & (2) & (3) & (4) & (5)\\ 
\hline \\[-1.8ex] 
 $<$ 0$^{\circ}$C & 0.044$^{***}$ & 0.033 & $-$0.036$^{*}$ & 0.014 & 0.059$^{***}$ \\ 
  & (0.013) & (0.028) & (0.020) & (0.018) & (0.011) \\    

 0$^{\circ}$C - 5$^{\circ}$C& 0.024$^{**}$ & 0.031$^{**}$ & $-$0.010 & 0.024 & 0.024$^{**}$ \\ 
  & (0.011) & (0.015) & (0.019) & (0.017) & (0.011) \\    

 5$^{\circ}$C - 10$^{\circ}$C  & 0.020$^{**}$ & 0.023$^{**}$ & $-$0.011 & 0.017 & 0.022$^{**}$ \\ 
  & (0.010) & (0.011) & (0.012) & (0.014) & (0.009) \\   

 10$^{\circ}$C - 15$^{\circ}$C & Ref. & Ref. & Ref. & Ref. & Ref.\\ 
 & & & & & \\ 

  15$^{\circ}$C - 20$^{\circ}$C & 0.005 & 0.008 & 0.009 & 0.023$^{*}$ & $-$0.006 \\ 
  & (0.009) & (0.006) & (0.010) & (0.013) & (0.009) \\    
  
 20$^{\circ}$C - 25$^{\circ}$C  & $-$0.001 & $-$0.004 & 0.014 & 0.003 & $-$0.004 \\ 
  & (0.017) & (0.013) & (0.016) & (0.019) & (0.012) \\ 

 $>$ 25$^{\circ}$C  & 0.013 & 0.045$^{**}$ & 0.035$^{**}$ & 0.053$^{***}$ & 0.010$^{*}$ \\ 
  & (0.013) & (0.020) & (0.017) & (0.017) & (0.005) \\ 
  & & & & & \\ 
Observations & 21,138 & 21,138 & 21,138 & 21,138 & 21,138 \\ 
Adjusted R$^{2}$  & 0.439 & 0.323 & 0.120 & 0.272 & 0.428 \\ 
\hline \\[-1.8ex] 
Counseling-center fixed effects & Yes & Yes & Yes & Yes & Yes \\ 
Year-by-month fixed effects & Yes & Yes & Yes & Yes & Yes \\
Environmental controls & Yes & Yes & Yes & Yes & Yes \\ 
Day-of-week fixed effects & Yes & Yes & Yes & Yes & Yes\\ 
Holiday fixed effects & Yes & Yes & Yes & Yes & Yes\\ 
\hline 
\hline \\[-1.8ex] 
\multicolumn{6}{l}{Notes: The dependent variable is the natural logarithm of the daily number of answered helpline calls. The sample period} \\ 
\multicolumn{6}{l}{is November 3, 2018 to February 29, 2020. The standard errors in parentheses are two-way clustered at the counseling} \\ 
\multicolumn{6}{l}{center and year-month level. Columns (1) and (2) present the results when the sample is split by gender. Columns (3) - (5) } \\
\multicolumn{6}{l}{present the results when the sample is split by age cohorts. $^{*}$p$<$0.1; $^{**}$p$<$0.05; $^{***}$p$<$0.01.} \\
\end{tabularx} 
\end{table}

\begin{table}[t!] 
\captionsetup{justification=centering,margin=1cm}
\centering 
\caption{Effect of average daily temperature on helpline call volume by main topic} 
\label{tab:results_top} 
\scriptsize
\begin{tabularx}{\textwidth}{lYYYYYY}
\\[-1.8ex]\hline 
\hline \\[-1.8ex]
 & \multicolumn{6}{c}{\textit{Dependent variable:}} 
 \\
 \cline{2-7} 
 \\[-1.8ex] & \multicolumn{6}{c}{helpline calls} \\
\\[-1.8ex]  & \multicolumn{6}{c}{main topic}\\
\cline{2-7} 
\\[-1.8ex] & physical & psychological & violence & social & liveli- & other\\ 
 & well-being & well-being &  & well-being & hood & \\ 

\\[-1.8ex] & (1) & (2) & (3) & (4) & (5) & (6)\\ 
\hline \\[-1.8ex] 
 $<$ 0$^{\circ}$C & 0.045$^{*}$ & 0.045$^{***}$ & $-$0.013 & 0.057$^{***}$ & 0.023 & 0.047$^{**}$ \\ 
  & (0.026) & (0.015) & (0.013) & (0.014) & (0.017) & (0.021) \\  

 0$^{\circ}$C - 5$^{\circ}$C & 0.008 & 0.021 & $-$0.024$^{**}$ & 0.037$^{***}$ & 0.033$^{**}$ & 0.019 \\ 
  & (0.021) & (0.014) & (0.012) & (0.010) & (0.017) & (0.018) \\

 5$^{\circ}$C - 10$^{\circ}$C & 0.001 & 0.019$^{**}$ & $-$0.029$^{**}$ & 0.033$^{***}$ & 0.023 & 0.022 \\ 
  & (0.019) & (0.008) & (0.012) & (0.007) & (0.014) & (0.015) \\ 

 10$^{\circ}$C - 15$^{\circ}$C & Ref. & Ref. & Ref. & Ref. & Ref.& Ref.\\ 
 & & & & & &\\ 

  15$^{\circ}$C - 20$^{\circ}$C & $-$0.016 & 0.020$^{*}$ & 0.010 & 0.006 & $-$0.015 & $-$0.005 \\ 
  & (0.014) & (0.011) & (0.013) & (0.009) & (0.019) & (0.016) \\ 
  
 20$^{\circ}$C - 25$^{\circ}$C & $-$0.012 & $-$0.003 & 0.023$^{*}$ & $-$0.013 & $-$0.010 & $-$0.002 \\ 
  & (0.021) & (0.019) & (0.014) & (0.017) & (0.029) & (0.022) \\

 $>$ 25$^{\circ}$C & $-$0.014 & 0.070$^{**}$ & 0.058$^{***}$ & $-$0.001 & 0.017 & 0.076 \\ 
  & (0.019) & (0.031) & (0.022) & (0.007) & (0.018) & (0.075) \\  
  & & & & & &\\ 
Observations & 21,138 & 21,138 & 21,138 & 21,138 & 21,138 & 21,138\\ 
Adjusted R$^{2}$  & 0.218 & 0.268 & 0.083 & 0.377 & 0.208 & 0.103 \\  
\hline \\[-1.8ex] 
Counseling-center fixed effects & Yes & Yes & Yes & Yes & Yes &Yes\\ 
Year-by-month fixed effects & Yes & Yes & Yes & Yes & Yes &Yes\\
Environmental controls & Yes & Yes & Yes & Yes & Yes &Yes\\ 
Day-of-week fixed effects & Yes & Yes & Yes & Yes & Yes&Yes\\ 
Holiday fixed effects & Yes & Yes & Yes & Yes & Yes&Yes\\ 
\hline 
\hline \\[-1.8ex] 
\multicolumn{7}{l}{Notes: The dependent variable is the natural logarithm of the daily number of answered helpline calls. The sample period} \\ 
\multicolumn{7}{l}{is November 3, 2018 to February 29, 2020. The standard errors in parentheses are two-way clustered at the counseling} \\ 
\multicolumn{7}{l}{ center and year-month level. Columns (1) - (6) present the results when the sample is split by main topic of the conver-}\\
\multicolumn{7}{l}{sation. $^{*}$p$<$0.1; $^{**}$p$<$0.05 $^{***}$p$<$0.01.}\\
\end{tabularx} 
\end{table}

\clearpage

\pagebreak

\begin{figure}[]
\centering
\captionsetup{justification=centering,margin=1cm}
\caption{Visibility of telephone counseling services, Panel (A): online search; Panel (B): information poster}
\label{fig:Telefonseelsorge_google}
\includegraphics[width=.9\textwidth]{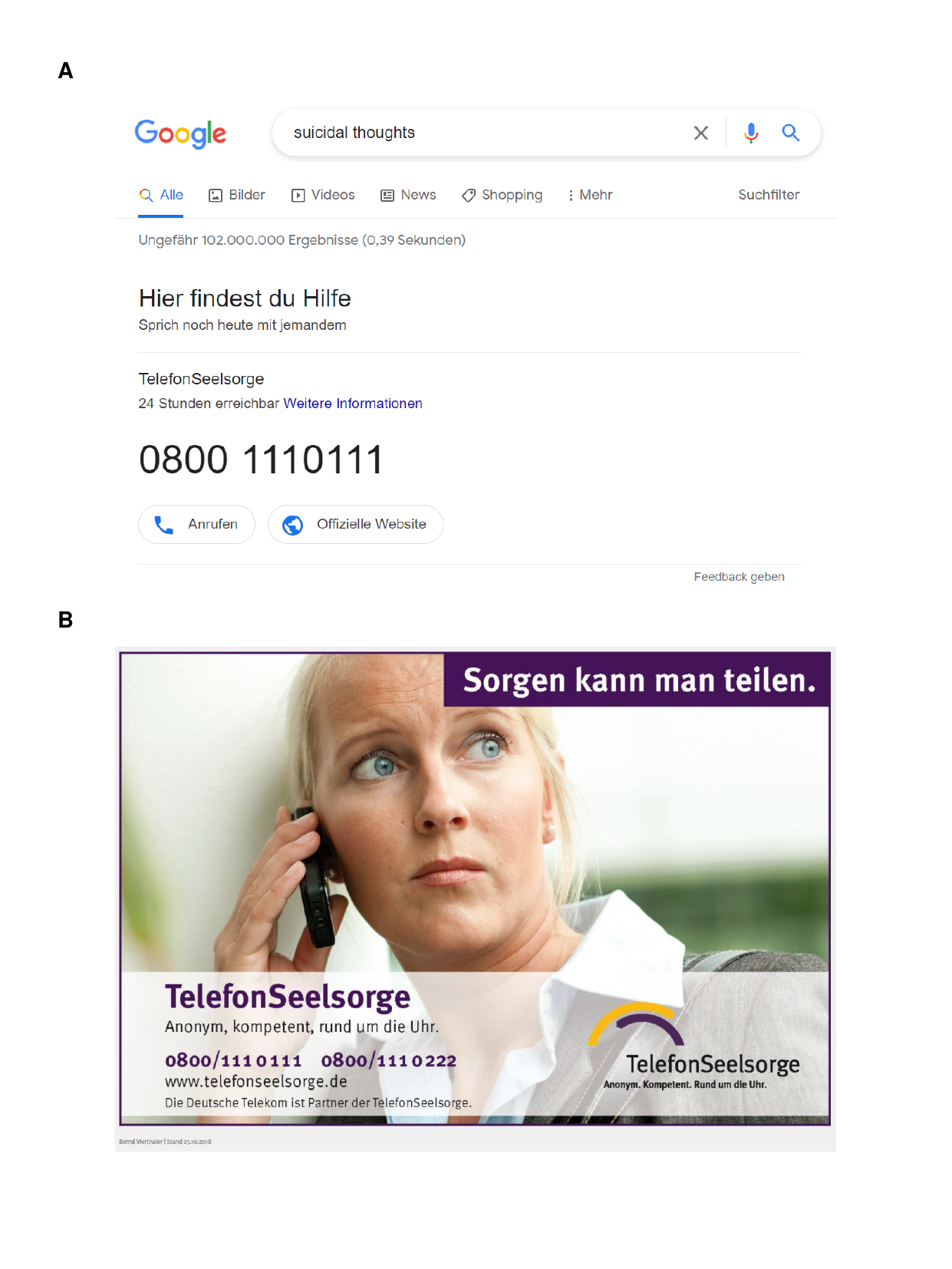}
\begin{minipage}{1\textwidth} 
{\scriptsize{Notes: Panel A: This figure presents a representative example of an online search for mental health concerns in Germany. The search term is 'suicidal thoughts'. The phone number of \textit{TelefonSeelsorge} is prominently placed on top of the search page. Above the phone number it says, 'You can find help here - Talk to someone today'. Panel B:  This figure presents a representative example of an information poster for telephone counseling services offered by \textit{TelefonSeelsorge}. In the top right corner it says, 'Worries can be shared.'. \par}}
\end{minipage}
\end{figure}

\begin{figure}[t]
\centering
\captionsetup{justification=centering,margin=1cm}
\caption{Catchment area of \textit{TelefonSeelsorge} Münster}
\label{fig:tel_munster}
\includegraphics[width=.9\textwidth]{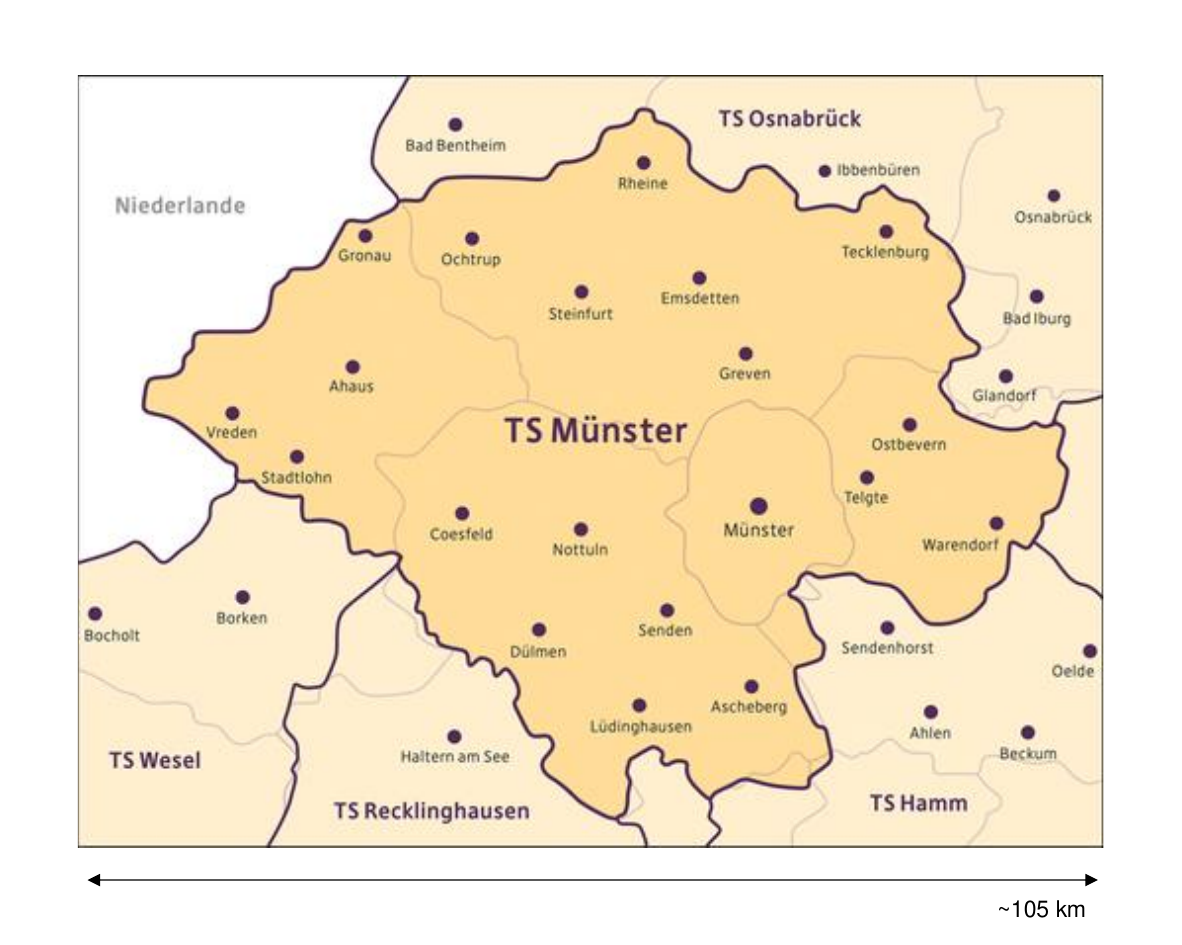}
\begin{minipage}{1\textwidth} 
{\scriptsize{This figure presents the catchment area of \textit{TelefonSeelsorge} Münster. Maps can be obtained from some of the crisis center's website. See \url{https://www.telefonseelsorge.de/unsere-stellen/} for a list of all locations and their websites. \par}}
\end{minipage}
\end{figure}

\begin{figure}[t]
\centering
\captionsetup{justification=centering,margin=1cm}
\caption{Distribution of the number of calls per day}
\label{fig:call_hist}
\includegraphics[width=.8\textwidth]{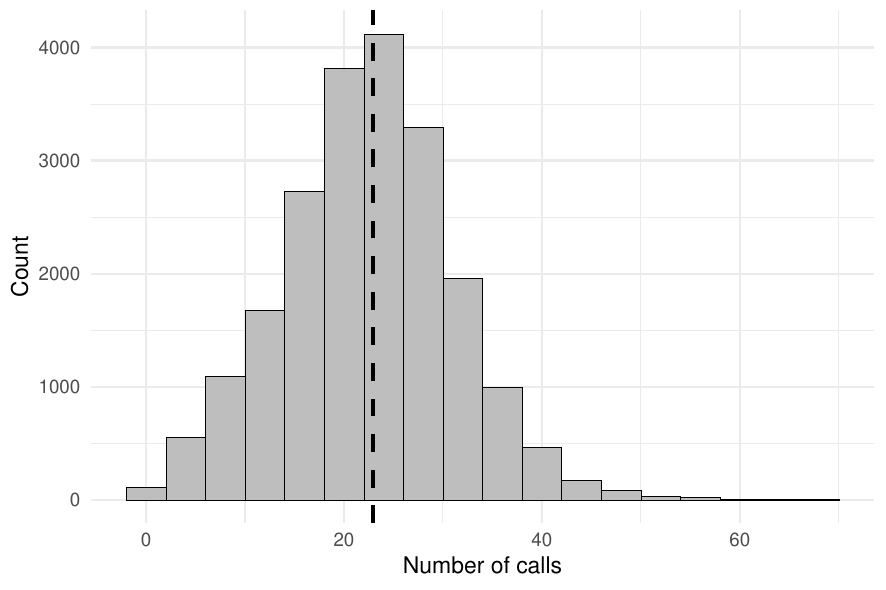}
\begin{minipage}{1\textwidth} 
{\scriptsize{Notes: This figure illustrates the distribution of the number of helpline calls answered at any given counseling-center-day in the study sample. The horizontal dashed line represents the sample mean (22.96).\par}}
\end{minipage}
\end{figure}

\begin{figure}[t]
\centering
\captionsetup{justification=centering,margin=1cm}
\caption{Distribution of helpline call duration}
\label{fig:dur_hist}
\includegraphics[width=.8\textwidth]{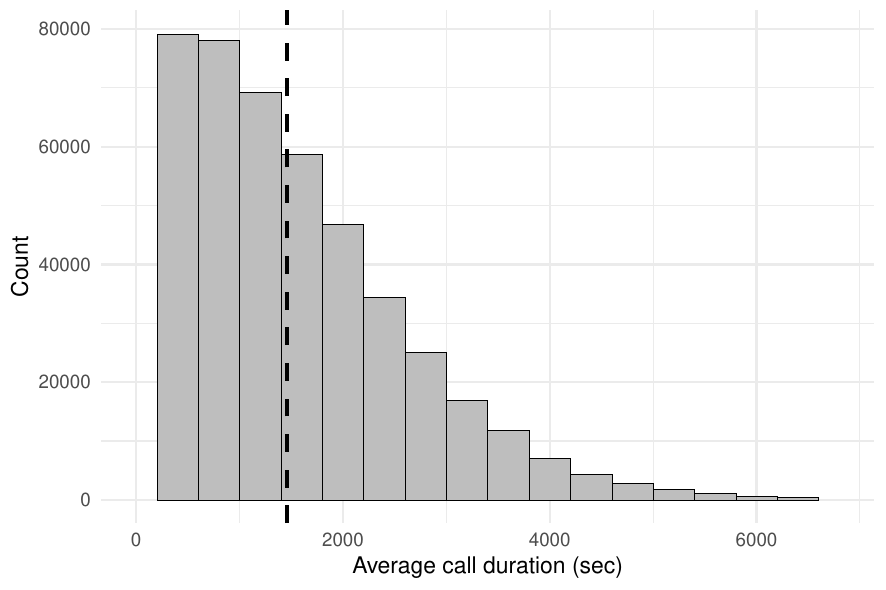}
\begin{minipage}{1\textwidth} 
{\scriptsize{Notes: This figure illustrates the distribution of the length of the calls answered at any given counseling center in the study sample. The horizontal dashed line represents the sample mean (1457.71 sec $\equiv$ \textrm{24 min 18 sec}).\par}}
\end{minipage}
\end{figure}

\begin{figure}[t]
\centering
\captionsetup{justification=centering,margin=1cm}
\caption{Distribution of concurrent environmental factors, Panel (A): precipitation; Panel (B): sunshine duration; Panel (C): surface wind speed; Panel (D): relative humidity; Panel (E): air pollution}
\label{fig:summary_envr}
\includegraphics[width=1\textwidth]{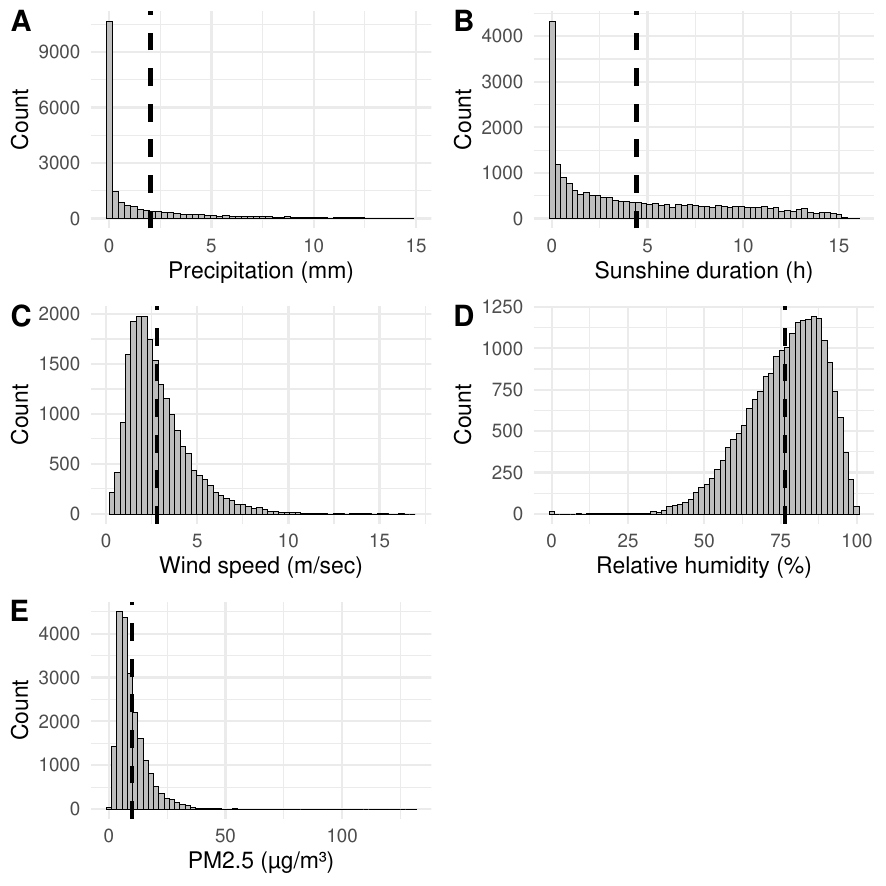}
\begin{minipage}{1\textwidth} 
{\scriptsize{Notes: This figure illustrates the distribution of realized daily average (A) precipitation, (B) sunshine duration, (C) wind speed, (D) humidity, and (E) air pollution at any given counseling-center-day in the study sample. The horizontal dashed line represents the respective sample mean.\par}}
\end{minipage}
\end{figure}

\begin{figure}[t]
\centering
\captionsetup{justification=centering,margin=1cm}
\caption{Simulation-based null distribution of regression coefficients for extreme temperature days, Panel (A): $<$0$^{\circ}$C ; Panel (B): $>$25$^{\circ}$C }
\label{fig:summary_simul}
\includegraphics[width=.8\textwidth]{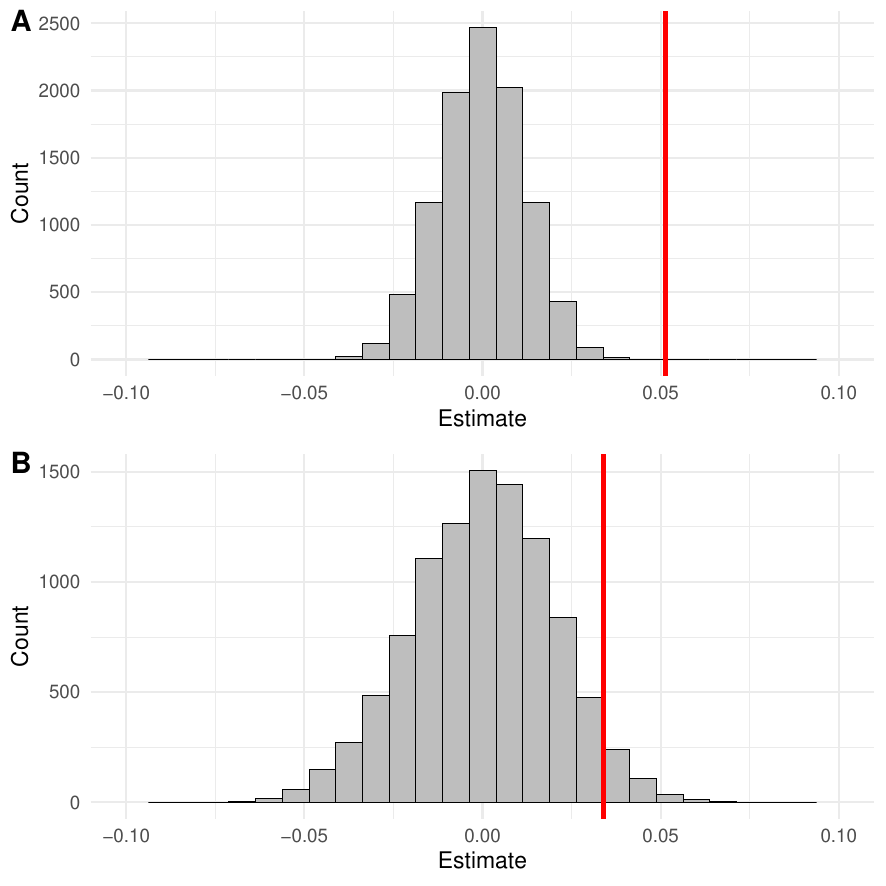}
\begin{minipage}{1\textwidth} 
{\scriptsize{Notes: This figure illustrates the simulation-based null distribution of the regression estimates for (A) $<$0$^{\circ}$C and (B) $>$25$^{\circ}$C. The horizontal red line represents the observed estimate, (A) 0.051 and (B) 0.034. Proportion of simulated coefficients that lie below the slope estimated on the real data is (A) 1 and (B) 0.960. Extreme temperature days are randomly permuted 10,000 times across counseling-center-day observations and the baseline model outlined in equation (\ref{eq:3}) is re-estimated for each of these 10,000 simulated datasets.\par}}
\end{minipage}
\end{figure}

\end{document}